\documentclass[a4paper,11pt]{article}

\usepackage{lineno,hyperref,url} 
\usepackage{amsmath,amsfonts,amssymb,amsthm,enumitem,stmaryrd,xfrac,nicefrac} 
\numberwithin{equation}{section} 
\usepackage{graphics,graphicx,subfigure} 
\usepackage{array,booktabs,diagbox,multirow,caption}  
\usepackage[most]{tcolorbox} 
\usepackage{soul}
\usepackage{tikz}
\usetikzlibrary{arrows}
\usepackage[textwidth=16cm]{geometry}

\newcommand{\footremember}[2]{%
    \footnote{#2}
    \newcounter{#1}
    \setcounter{#1}{\value{footnote}}%
}
\newcommand{\footrecall}[1]{%
    \footnotemark[\value{#1}]%
} 

\usepackage[round]{natbib}
\newcommand{\bs}{\boldsymbol} 
\newcommand{\mb}{\mathbb} 
\newcommand{\mc}{\mathcal} 
\newcommand{\mr}{\mathrm} 
\newcommand{\pd}{\partial}
\newcommand{\wt}{\widetilde} 
\newcommand{\wh}{\widehat} 
\newcommand{\ol}{\overline}
\newcommand{\sub}{\scriptscriptstyle} 
\newcommand{\beq}{\begin{equation}}
\newcommand{\eeq}{\end{equation}}
\newcommand{\beqs}{\begin{subequations}}
\newcommand{\eeqs}{\end{subequations}}
\newcommand{\benum}{\begin{enumerate}[label=(\roman*)]}
\newcommand{\eenum}{\end{enumerate}}

\newcommand\real{\mbox{Re}} 
\newcommand\imag{\mbox{Im}} 

\newcommand{\dom}{\mc{D}} 

\newcommand{\df}{\mr{d}} 
\newcommand{\sto}{\mr{\mb{D}}} 
\newcommand{\ialf}{\text{\scriptsize$\frac{i}{2}$}} 
\newcommand{\alf}{\text{\scriptsize$\frac{1}{2}$}} 

\newcommand{\kerl}{\breve{\sigma}} 
 
\newcommand{\bdot}{\bs{\cdot}} 
\newcommand{\grad}{\bs{\nabla}} 
\newcommand{\gradp}{\grad^{\perp}} 
\newcommand{\bdiv}{\grad \bdot} 

\newcommand{\curl}{\grad \times} 
\newcommand{\adv}{\bdot \grad} 
\newcommand{\tp}{^{\sub T}} 
\newcommand{\tc}{^{\dagger}}
\newcommand{\Exp}{\mb{E}} 

\newcommand{\dt}{\df t} 
 
\newcommand{\bsig}{\bs{\sigma}_t} 
\newcommand{\bkerl}{\breve{\bs{\sigma}}} 
\newcommand{\bxi}{\bs{\xi}} 

\newcommand{\B}{\bs{B}} 
\newcommand{\noi}{\bsig \df \B_t}
\newcommand{\rsf}{\varphi \df B_t}

\newcommand{\bI}{\bs{\mr{I}}} 
\newcommand{\x}{\bs{x}}
\newcommand{\y}{\bs{y}}

\newcommand{\q}{\bs{q}}
\newcommand{\bk}{\bs{k}}
\newcommand{\bks}{\bk_{\sigma}}
\newcommand{\bu}{\bs{u}}
\newcommand{\bL}{\bs{\mr{L}}}
\newcommand{\bv}{\bs{v}}

\newcommand{\ba}{\bs{a}}

\newcommand{\dx}{\df \x} 
\newcommand{\dy}{\df \y}

\usepackage{comment,color}

\newcommand{\delm}[1]{\ifmmode\text{\sout{\ensuremath{#1}}}\else\sout{#1}\fi} 

\title{\vspace{-2.75cm} Linear wave solutions of a stochastic shallow water model}

\author{%
  Etienne M\'emin\footnote{Corresponding author}\footremember{inria}{ODYSSEY Team, Centre Inria de l'Universit\'e de Rennes, France}%
  \and Long Li\footrecall{inria}%
  \and No\'{e} Lahaye\footrecall{inria}%
  \and Gilles Tissot\footrecall{inria}%
  \and Bertrand Chapron\footremember{ifremer}{Laboratoire d’Oc\'eanographie Physique et Spatiale, Ifremer, Plouzan\'e, France}%
  }

\date{}

\begin{document}

\maketitle

\begin{abstract}
In this paper, we investigate the wave solutions of a stochastic rotating shallow water model. This approximate  model provides an interesting simple description of the interplay between waves and random forcing ensuing either from the wind or coming as the feedback of the ocean on the atmosphere and leading in a very fast way to the selection  of some wavelength. This interwoven, yet simple, mechanism explains the emergence of typical wavelength associated with near inertial waves. Ensemble-mean waves that are not in phase with the random forcing are damped at an exponential rate, whose magnitude  depends on the random forcing variance. Geostrophic adjustment is also interpreted  as a statistical homogenization process in which, in order to conserve potential vorticity, the small-scale component tends to align to the velocity fields to form a statistically homogeneous random field. 
\end{abstract}
\vspace{0.5cm}


\section{Introduction}
Oceanic global circulation currents show a predominance of near inertial waves (NIW) in their spectrum. These waves result from the repeated forcing of atmospheric winds over the globe together with the influence of global earth rotation at Coriolis frequency. Besides, large-scale numerical representations of oceanic circulation require to introduce subgrid scale models that inescapably damp in the long run velocity fields and wave solutions. In such models high frequency waves are obviously completely smoothed out. But even large scale structures might be also affected if no particular attention is paid to the subgrid model design. 

In the last year, there has been an increasing effort to devise stochastic parameterizations for large-scale flows \cite{Berner-17,Franzke2015,Gottwald2017}. The motivations come from the failure of classical subgrid models to represent accurately the effect of the flow state variables at the unresolved scales and the necessity to provide reliable and computationally efficient models at the climatic scale. Uncertainty quantification, ensemble methods for forecasting and data assimilation are also prevailing, and the Bayes principle on which they are built leads naturally to considering stochastic dynamics. 

Stochastic dynamics can hardly be devised on {\em ad hoc}  grounds if one wants to provide generic and flexible systems. Control of the variance's growth and the respect of the  physical properties of the underlying turbulent flows is of the utmost importance to define stable and physically relevant systems, in which the unresolved variables or the different physical and numerical approximations performed are faithfully represented. Two main schemes have been recently proposed to that purpose in the literature \cite{Holm-15,Memin14}. The first one is a geometric framework relying on a Hamiltonian formulation, whereas the second one -- referred to as modelling under location uncertainty (LU) -- is based on Newton's principles and a stochastic formulation of the Reynolds transport theorem. Both schemes have been analyzed and numerically experimented for several  geophysical models and configurations \cite{Bauer-et-al-JPO-20,Brecht2021,Chapron-18,Cotter-MMS-19,Li-James-2023,Resseguier2020arcme,Resseguier2020}. Surface waves and linear models have been proposed  in the LU framework \cite{Dinvay-Memin-22,Tissot-JFM-21,Tissot-PRF-23}. 

Nevertheless, the analysis of basic geostrophic mechanisms such as geostrophic adjustment or the form of basic wave solutions in linearized simple systems has not been investigated so far. This is the objective of this work. We will in particular analyze a stochastic rotating shallow water model recently proposed in \cite{Brecht2021,Lang2022}. The focus will be on wave solutions of this system. We will show in particular that ensemble-mean waves ensuing from a random forcing at given frequencies are preserved whereas the others are very quickly damped. The stochastic model will allow us to propose a very simple mechanism leading to the emergence and conservation of NIW in atmosphere and ocean dynamics as a coupled self entrainment process.

The paper is organized as follows. In a first section we briefly recall the proposed stochastic model -- rotating shallow water under location uncertainty (RSW-LU). We then identify a stationary solution of this stochastic system. We next analyze and illustrate the plane wave solutions associated to the linearized system under some specific noises. Finally, we provide a picture of the geostrophic adjustment process associated to the random system. 


\section{Review of RSW-LU}

Let us first recall the stochastic transport operator introduced in the LU framework \cite{Bauer-et-al-JPO-20,Li-Memin-Tissot-22,Li-James-2023,Resseguier-GAFD-I-17}:
\beqs\label{eqs:sto-transp}
\beq
\sto_t \Theta := \df_t \Theta + \Big( \big( \bu - \alf \bdiv \ba \big)\, \dt + \noi \Big) \adv \Theta - \alf \bdiv \big( \ba \grad \Theta \big)\, \dt = 0,
\eeq
where the tracer, $\Theta$, is a stochastic process with an extensive property (e.g. temperature, salinity, buoyancy), $\df_t \Theta (\x) := \Theta (\x, t+\delta t) - \Theta (\x, t)$ stands for the time-increment of $\Theta$ at a fixed point $\x$ with $\delta t$ an infinitesimal time variation, $\bu$ denotes the time-smooth resolved velocity that is both spatially and temporally correlated, $\noi$ stands for the highly oscillating unresolved noise component, assumed in the present study to be divergence-free, with its quadratic variation \cite{DaPrato} denoted by $\ba$ and $\alf \bdiv \ba$ is the so-called It\^o-Stokes drift \cite{Bauer-et-al-JPO-20} ensuing from the inhomogeneity of the noise. The mathematical definitions of the noise and its quadratic variation are given by
\beq\label{eq:noise}
\bsig  \df \B_t (\x)= \int_{\dom} \bkerl (\x, \y, t) \df \B_t (\y)\, \dy = \sum_{n \in \mb{N}} \lambda_n^{1/2} (t) \bxi_n (\x, t) \df \beta_t^n,
\eeq
\beq\label{eq:a}
\ba (\x, t) = \int_{\dom} \bkerl (\x, \y, t) \bkerl\tc (\y, \x, t)\, \dy = \sum_{n \in \mb{N}} \lambda_n (t) \big( \bxi_n \bxi_n\tc \big) (\x, t), 
\eeq
\eeqs
where $\bsig$ is an integral operator defined on the Hilbert space $(L^2 (\dom))^d$ with a bounded spatial domain $\dom \subset \mb{C}^d$ ($d = $ 2 or 3), 
$\bkerl = (\kerl_{ij})_{i,j=1,\ldots,d}$ is a spatially and temporally bounded matrix kernel of  $\bsig$, $\lambda_n$ and  $\bxi_n$ are the eigenfunctions and eigenvalues of the composite operator $\bsig \bsig^*$ ($\bsig^*$ denotes the adjoint of $\bsig$), $\bullet\tc$ stands for transpose-conjugate operation, $\B_t$ is a cylindrical Wiener process \cite{DaPrato} and $\beta_t^n$ are independent (one-dimensional) standard Brownian motions.

Under the stochastic transport notations \eqref{eqs:sto-transp}, the governing equations of the energy-preserved RSW-LU system \cite{Brecht2021} read
\beqs\label{eqs:rswlu}
\begin{align}
&{(\text{\em Conservation of momentum})}\nonumber\\
&\sto_t \bu + f_0 \bu^\perp\, \dt = - g \grad \eta\, \dt, \label{eq:moment}\\
&{(\text{\em Conservation of mass})}\nonumber\\
&\sto_t h + h \bdiv \bu\, \dt = 0, \label{eq:mass}\\
&{(\text{\em  Incompressible constraints})}\nonumber\\
&\noi = \gradp \rsf,\ \quad \bdiv \bdiv \ba = 0, \label{eq:incomp}\\
&{(\text{Conservation of energy})}\nonumber\\
&\df_t \int_{\dom} \alf \rho \big( h |\bu|^2 + g h^2 \big)\, \dx = 0,  \label{eq:energy}
\end{align}
\eeqs
where $\bu = [u, v]\tp$ denotes the two-dimensional horizontal velocity with $\bu^\perp = [-v, u]\tp$, $\grad = [\pd_x, \pd_y]\tp$ stands for the horizontal gradient with $\grad^\perp = [-\pd_y, \pd_x]\tp$, $h (\x, t) = H + \eta (\x, t)$ is the water thickness with $H$ the  flat bottom height and $\eta$ the free-surface position, 
$f_0$ is a constant approximation of the Coriolis parameter,
$g$ is the gravity constant, $\rho$ is the water density and $\rsf$ denotes a random (scalar) stream function defined in a similar way as in \eqref{eq:noise}. As shown in \cite{Brecht2021}, the incompressible conditions \eqref{eq:incomp} for both noise and It\^o-Stokes drift ensure the path-wise conservation of the total energy \eqref{eq:energy}. We remark that the analytical properties for the RSW-LU system have been investigated in \cite{Lang-et-al-23} and some numerical applications of \eqref{eqs:rswlu} have been performed in \cite{Brecht2021}.


\section{Stationary solution}

We focus now on stationnary solutions associated to the previous system. To that end, neglecting the time increments $\df_t \bu$ and $\df_t h$ in \eqref{eq:moment}--\eqref{eq:mass} and recalling that $\grad \eta = \grad h$ due to the flat bottom assumption, we obtain 
\beqs\label{eqs:rswlu-sta}
\beq\label{eq:moment-sta}
\Big( \big( \bu - \alf \bdiv \ba \big) \adv \bu - \alf \bdiv (\ba \grad \bu) + f_0 \bu^\perp + g \grad h \Big)\,\dt + \noi \adv \bu = 0,  
\eeq
\beq\label{eq:mass-sta}
\Big( \big( \bu - \alf \bdiv \ba \big) \adv h - \alf \bdiv (\ba \grad h) + h \bdiv \bu \Big)\,\dt + \noi \adv h = 0,  
\eeq
\beq\label{eq:incomp-sta}
\noi = \gradp \rsf, \quad \ba = \gradp \varphi (\gradp \varphi)\tp.
\eeq
\eeqs
From Bichteler-Dellacherie decomposition of a semi-martingale, the martingale terms (i.e. the Brownian terms) and the finite variation terms (i.e. the differentiable terms) can be safely separated. Decomposing in such way the mass equation~\eqref{eq:mass-sta} and considering the corresponding martingale part leads to $\noi \adv h =0$. The random surface height gradient is aligned with the large-scale surface height gradient  
\begin{equation}\label{dpt}
\noi = \alpha \gradp h,
\end{equation}
The noise being divergence free yields that $\alpha$ is constant  along the level sets of $h$:
\begin{equation}
    \grad \alpha\bdot \gradp h= 0. 
\end{equation} 
It is hence necessary of the form $\alpha = \phi(h) +C$, with $C$ a constant, and $\phi$ a differentiable function.
For such a  noise the variance tensor is given by 
\begin{equation}
\ba = \alpha^2 \gradp h  \gradp h\tp,
\end{equation}
while the ISD reads 
\begin{equation}
\label{ISD-ADV}
\bdiv\ba = \alpha^2 \bdiv (\gradp h {\gradp}h\tp) =\alpha^2  (\gradp h \bdot \grad) \gradp h,
\end{equation}
 and we notice that $\ba\grad h =0$.
The mass equation then boils down to 
\beqs
\beq
\big( \bu - \alf \bdiv \ba \big) \adv h + h \bdiv \bu  = 0.
\eeq
In the same way, we get from a semi-martingale decomposition of \eqref{eq:moment-sta} that 
\beq
\label{nab-perp-null-space}
\noi \adv \bu = 0, 
\eeq
which yields 
\beq
\big( \bu - \alf \bdiv \ba \big) \adv \bu + f_0 \bu^\perp + g \grad h = 0.
\eeq
\eeqs
At this point it is interesting to interpret the random streamfunction in terms of nondimensional units to infer the importance of the It\^o-Stokes drift (ISD) in comparison to geostrophic flows. To that end, we assume that $\rsf \sim  \sqrt{\epsilon T}(g H / f_0) \sqrt{T}$, where $T$ denotes the characteristic time scale and $\epsilon T$ stands for a small-scale decorrelation time with $\epsilon$ the strength of uncertainty (the greater $\epsilon$ the stronger the noise is). From definition \eqref{eq:incomp-sta}, the noise's quadratic variation scales then as $\ba \sim \epsilon T (g H)^2 / (f_0 L)^2$ with $L$ the characteristic length scale. As a consequence, the ratio of the ISD advection term with the geostrophic gradient pressure scales as
\beq
\frac{(\bdiv \ba) \adv \bu}{g \grad h} \sim \epsilon \frac{g H}{f_0^2 L^2} = \epsilon \frac{L_d^2}{L^2} = \epsilon \rm{B_u},
\eeq
where $L_d = \sqrt{g H} / f_0$ is the Rossby deformation radius, and $\rm{B_u}$ denotes the Burger number, which stands for the ratio between vertical density stratification and the earth's rotation in the horizontal ($\rm{B_u}= (NH/\Omega L)^2 =  (L_d/L)^2$). 
If the ISD advection has the same importance as the gradient pressure term, then we must have $\sqrt{\epsilon} = L / L_d$. This means that when the scale of motions is greater than the deformation radius, the rotation will dominate and the noise must be important to have the ISD playing a role. At the opposite, when the scale of motions is smaller than the Rossby radius, the small-scale flow component does not need to be important to be as significant as the gradient pressure term. From this point of view, the deformation radius can be interpreted as the limiting scale under which the statistically modified advection due to the inhomogeneity of the small-scale component plays a role.   

In the following, we will assume to be at a scale much larger than the deformation radius, so that the action of the ISD becomes negligible. In that case, the  stationary system finally simplifies as
\beq\label{eq:rsw-sta-final}
(\bu \adv) \bu + f_0 \bu^\perp + g \grad h = \bdiv (\bu h) = 0.
\eeq
This system is structurally the same as the nonlinear stationary system, at the exception of a scaling constraint on the ISD, which makes negligible the nonlinear advection term. As a matter of fact, noticing that for $\bu \propto \gradp h$, the advection term corresponds to the ISD \eqref{ISD-ADV}, it follows that the geostrophic balance, $\bu = - (g / f_0) \gradp h$, is a stationary solution of such system for a null ISD. We show below it is the only non-trivial stationary solution of such system.

Let us write the velocity as a superposition of the geostrophic component and an ageostrophic component:
\begin{equation}
    \bu= -\frac{g}{f_0}\gradp h + \bv,
\end{equation}
where $\bv$ is defined through the Helmholtz decomposition from a potential function, $\Phi$,  and a stream function $\Psi$ that both depend on the surface elevation:   
\begin{align}
     \bv &= \gradp \Psi(h) + \grad \Phi(h),\\
     &= \gradp h \;\Psi'(h) + \grad h \;\Phi'(h).
\end{align}
    As $\gradp h$ belongs to the null space of the velocity gradient \eqref{nab-perp-null-space},  $\grad h$ either belongs also to the null space  or it is an eigenvector of the velocity gradient tensor $(\grad \bu)$. 
From the momentum equation \eqref{eq:rsw-sta-final}  we have
   \[
   \bigl( \grad h \Phi'(h) \bigr)\bdot \grad \bv = -f \bv^\perp.
   \]
   Then if $\grad h$ belongs to the null space of the velocity gradient, $\bv$ directly cancels out. If is an eigenvector of the velocity gradient with eigenvalue $\lambda$, the above equation reads 
\[
    \lambda \grad h \Phi'(h) = - f \bigl(\grad h \Psi'(h) + \gradp h \Phi'(h)\bigr),
\]
which implies $\Phi'(h) = \Psi'(h)= 0$ and hence $\bv=0$. Physically, we see hence that the considered nonlinear system admits a stationary solution for a negligible ISD. Let us note that no such constraint is available in the deterministic setting. In the following we will interpret this stationary solution in the linearized stochastic shallow water system in terms of wave solutions and geostrophic adjustment. 


\section{Stochastic rotating shallow water waves}\label{sec:sw-wave-lu}

In order to look at the different wave solutions associated to the stochastic shallow water system \eqref{eqs:rswlu}, we proceed in the same way as in the deterministic case \cite{Majda2003,Vallis-17}. In particular, we assume that the noise's structure in \eqref{eq:noise} is independent of the resolved prognostic variable $\bu$. In that case, the associated linearized system of \eqref{eqs:rswlu} reads 
\beqs\label{eqs:rswlu-linear}
\beq\label{eq:moment-linear}
\df_t \bu + \Big( f_0 \bu^\perp + g \grad \eta - \alf \bdiv (\ba \grad \bu) \Big)\, \dt + \noi \adv \bu = 0,
\eeq
\beq
\df_t \eta + \Big( H \bdiv \bu - \alf \bdiv (\ba \grad \eta) \Big)\, \dt + \noi \adv \eta = 0, 
\eeq
\beq
\bdiv \noi = \bdiv \ba = 0,
\eeq
\beq
\df_t \int_{\dom} \alf \rho \big( H |\bu|^2 + g (H + \eta)^2 \big)\, \dx = 0.
\label{total-energy-linear-sys}
\eeq
\eeqs
It can be checked that this system conserves the total energy \eqref{total-energy-linear-sys} in the same way as the initial nonlinear stochastic system \eqref{eqs:rswlu} does. For noise defined through Hilbert-Schmidt correlation tensor this system admits a mild solution \cite{DaPrato}. Existence of strong solution could also be shown from the nonlinear system \cite{Lang-et-al-23}. In order to build some simple analytical solutions of this linearized system and to better understand the physical behaviours of the resulting waves, only very specific noise models will be considered in the following. We first build the ensemble-mean wave solutions under homogeneous noise, then investigate the path-wise solutions under constant noise and under homogeneous noise but with very smooth structures.

\subsection{Ensemble-mean waves under homogeneous noise}
\label{Sec-Mean}
Let us first recall the definition of the homogeneous and incompressible noise: 
\beqs
\begin{align}\label{eq:noise-homog}
\bsig (\x) \df \B_t &= \int_{\dom} \gradp \varphi (\x - \y) \df \B_t (\y)\, \dy \nonumber\\
&= \sum_{m} i \bk_m^\perp \wh{\varphi} (\bk_m) \df \beta_t (\bk_m) \exp (i \bk_m \bdot \x),
\end{align}
\beq
\ba = \sum_{m} |\wh{\varphi} (\bk_m)|^2 \bk_m^\perp (\bk_m^\perp)\tp,
\eeq
\eeqs
where $i$ denotes the imaginary unit, $\bk_m = [k_m, \ell_m]\tp \in \mb{R}^2$ is the $m$-th wavenumber vector, $\wh{\bullet}$ stands for Fourier transform (in space) coefficient and $\beta_t \in \mb{C}$ are independent Brownian motions satisfying $\beta_t (-\bk_m) = \ol{\beta_t (\bk_m)}$ with $\real \{\beta_t\}$ and $\imag \{\beta_t\}$ be independent. This noise is  homogeneous, and thus associated to a constant matrix $\ba$. Its ISD is null and fits naturally the condition on the stationary solution found in the previous section. 

For sake of simplicity, we assume hereafter that the noise is defined by only one Fourier mode (associated to a wavenumber $\bks$) combined with its complex conjugate, namely
\beq\label{eq:a-const}
\bsig (\x) \df \B_t = \real \big\{ i \bks^\perp \alpha \exp (i \bks \bdot \x)\, \df \beta_t \big\},\ \quad \ba = |\alpha|^2 \bks^\perp (\bks^\perp)\tp,
\eeq
where $\alpha = \wh{\varphi} (\bks)$ is assumed to be deterministic and real. 
This monochromatic noise can be directly extended to a multi-scale noise model \eqref{eq:noise-homog}. Results with a multiscale version of the noise  will be shown in the numerical section.

Taking now the expectation ($\Exp$) of the linearized random system \eqref{eqs:rswlu-linear}, we have
\beqs\label{eqs:rswlu-linear-mean}
\beq
\pd_t \Exp \big[ \bu \big] + f_0 \Exp \big[ \bu \big]^\perp + g \grad \Exp \big[ \eta \big] - \alf \bdiv (\ba \grad \Exp \big[ \bu \big]) = 0,
\eeq
\beq
\pd_t \Exp \big[ \eta \big] + H \bdiv \Exp \big[ \bu \big] - \alf \bdiv (\ba \grad \Exp \big[ \eta \big]) = 0, 
\eeq
\eeqs
where $\pd_t$ denotes the partial time derivative. In order to infer the mean wave solutions, we look for a deterministic ansatz of the form
\beq\label{eq:ansatz-mean}
\Exp \big[ \wt{\q} \big]\, (\x, t) = \wh{\q}_0\, \exp \big( i (\bk \bdot \x - \omega t) \big),
\eeq
where $\q = [\bu, \eta]\tp = \real \big\{ \wt{\q} \big\}$ is a compact notation for the prognostic variables of the RSW-LU \eqref{eqs:rswlu-linear}, $\wh{\q}_0$ is the initial constant vector (also assumed to be deterministic) and $\omega$ is the time-frequency. We remark that $\Exp[\wt{\q}] = \wt{\Exp[\q]}$ due to the commutativity of expectation with linear operators. 

Injecting next the previous ansatz together with the constant matrix $\ba$ \eqref{eq:a-const} into the system \eqref{eqs:rswlu-linear-mean}, we get $\bL \wh{\q}_0 = 0$ with
\beq\label{eq:L-mean}
\bL = \Big( -i \omega + \alf |\alpha|^2 (\bk_\sigma \times \bk)^2 \Big)\, \bI_3 +
\begin{bmatrix} 
0 & -f_0 & i g k \\ 
f_0 & 0 & i g \ell \\ 
i H k & i H \ell & 0 
\end{bmatrix},     
\eeq
where $\bI_3$ stands for the $3 \times 3$ identity matrix, $\bk_\sigma \times \bk = k_\sigma \ell - \ell_\sigma k$ denotes the cross product between the wavenumber vectors $\bk_{\sigma}$ and $\bk$ with $|\bk_\sigma \times \bk| = |\bk_\sigma| |\bk| \sin (\theta)$ and $\theta$ is the angle between them. 

As usual, the dispersion relations are then given by the solution of $\det (\bL) = 0$, namely
\beq\label{eq:omega-mean}
\omega = - \ialf |\alpha|^2 (\bk_\sigma \times \bk)^2,\ \quad \omega = - \ialf |\alpha|^2 (\bk_\sigma \times \bk)^2 \pm \sqrt{g H |\bk|^2 + f_0^2}.
\eeq
We realize immediately that when the noise has the same direction as the initial wave (i.e. $\bks \times \bk = 0$), these two frequencies correspond to the steady and Poincar\'e (inertia-gravity) waves of the deterministic system \cite{Majda2003,Vallis-17}. Conversely, when the noise is not aligned to the initial wave, then the term $- \ialf |\alpha|^2 (\bk_\sigma \times \bk)^2$ leads to a damping of the ensemble mean  wave. We next construct the mean plane wave solutions associated to the frequencies \eqref{eq:omega-mean}.

\subsubsection{Mean Poincar\'e waves}

With the value of the last two (opposite) frequencies in \eqref{eq:omega-mean} and the associated eigenvector of $\bL$ \eqref{eq:L-mean}, one obtains the following polarization relations:
\beq
\Exp \big[ \wt{\q} \big] = 
\begin{bmatrix} 
\frac{\omega k + i f_0 \ell}{H |\bk|^2} \\ 
\frac{\omega \ell - i f_0 k}{H |\bk|^2} \\ 
1 
\end{bmatrix} 
\wh{\eta}_0 \exp \big( i (\bk \bdot \x - \omega t) \big) \exp \big( - \alf |\alpha|^2 (\bk_\sigma \times \bk)^2 t \big).
\eeq
Taking the real part, we  finally deduce the ensemble-mean wave solution:
\beqs\label{eqs:mean-wave}
\beq    
\Exp \big[ \eta \big] = \wh{\eta}_0 \cos \big( \bk \bdot \x - \omega t \big) \exp \big( -\alf |\alpha|^2 (\bk_\sigma \times \bk)^2 t \big),
\eeq    
\beq
\Exp \big[ \bu \big] = \Exp \big[ u_\parallel \big] \frac{\bk}{|\bk|} + \Exp \big[ u_\perp \big] \frac{\bk^\perp}{|\bk|},
\eeq
\beq
\Exp \big[ u_\parallel \big] (\x, t) = \frac{\wh{\eta}_0 \omega}{H |\bk|} \cos \big( \bk \bdot \x - \omega t \big) \exp \big( - \alf |\alpha|^2 (\bk_\sigma \times \bk)^2 t \big),
\eeq
\beq
\Exp \big[ u_\perp \big] (\x, t) = \frac{\wh{\eta}_0 f_0}{H |\bk|} \sin \big( \bk \bdot \x - \omega t \big) \exp \big( - \alf |\alpha|^2 (\bk_\sigma \times \bk)^2 t \big),
\eeq    
\eeqs
where the component  $\Exp[u_\parallel]$ is associated to the mean pressure waves that depends on surface elevation mean, whereas the latter component $\Exp[u_\perp]$ is associated to mean vorticity waves that are initiated by rotation. Note that in the short waves limit with $\|\bk\|^2 \gg 1 / L_d^2$, the mean Poincar\'e wave corresponds to the inertia-gravity wave of a shallow water system without rotation. The damping term $\exp \big( -\alf |\alpha|^2 (\bk_\sigma \times \bk)^2 t \big)$ depends on the noise's wavelength and variance. This term is zero when the noise and the wave are colinear (i.e. the angle $\theta$ between $\bk_\sigma$ and $\bk$ is zero). For high noise magnitude (and $\theta \neq 0$), the damping occurs in a very fast way. In the long wave limit with $\|\bk\|^2 \ll 1 / L_d^2$, the frequency approaches the Coriolis frequency and the damping term is much less important unless the noise amplitude is very high. Nevertheless, the exponential damping in time remains when the noise and the waves are not aligned. 

\subsubsection{Mean geostrophic mode}

The polarization for the eigenvalue $\omega = -\alf |\alpha|^2 (\bk_\sigma \times \bk)^2$ reads
\beq
\Exp \big[ \wt{\q} \big] (\x, t) = 
\begin{bmatrix} 
-i \frac{g}{f_0} \ell \\ 
 i \frac{g}{f_0} k \\ 
1 
\end{bmatrix} 
\wh{\eta}_0 \exp \big( i \bk \bdot \x \big) \exp \big( - \alf |\alpha|^2 (\bk_\sigma \times \bk)^2 t \big).
\eeq  
The ensemble-mean of the wave solutions are given by
\beqs\label{eqs:steady-mean}
\beq
\Exp \big[ \eta \big] = f_0 \wh{\eta}_0 \cos \big( \bk \bdot \x\big) \exp \big( -\alf |\alpha|^2 (\bk_\sigma \times \bk)^2 t \big),
\eeq
\beq
\Exp \big[ \bu \big] = - g \wh{\eta}_0 \bk^\perp \sin \big( \bk \bdot \x \big) \exp \big( -\alf |\alpha|^2 (\bk_\sigma \times \bk)^2 t \big).
\eeq
\eeqs
This is a pure vorticity wave. When the noise and the wave are aligned, it corresponds to a steady wave in  geostrophic balance, which is also, as we saw, a stationary solution of the nonlinear stochastic system associated to a divergence-free quadratic variation \eqref{eq:rsw-sta-final}. We next look at the path-wise wave solutions under specific noise.

\subsection{Path-wise waves under constant noise}

As an initial informative example, we first assume that the noise is constant in space using a zeroth order approximation of the Fourier mode $\exp (i \bks \bdot \x)$ in \eqref{eq:a-const}. It can be expressed as
\beq\label{eq:noise-const}
\noi = \real \big\{ i \alpha \bks^\perp \df \beta_t \big\},\ \quad \ba = \alpha^2 \bks^\perp (\bks^\perp)\tp.
\eeq
In order to infer wave solutions, we look for stochastic ansatz of the form
\beqs\label{eqs:ansatz-const}
\beq
\wt{\q}\, (\x, t) = \wh{\q}_0\, \exp \Big( i \big( \bk \bdot \x - \omega t - \real \{ i \gamma \beta_t \} \big) \Big), 
\eeq
where 
$\gamma$ is of unit s$^{-1/2}$. This ansatz has been shown to be a solution of a linear stochastic water waves for constant noise in \cite{Dinvay-Memin-22}.
Applying now the It\^o formula \cite{DaPrato} for the deterministic function $(t, z) \mapsto \wh{\q}_0\, \exp \big( i (c - \omega t - z) \big)$ composed with the random process $\real \{i \gamma \beta_t\}$, we deduce
\beq\label{eq:dt-ansatz}
\df_t \wt{\q} = - \Big( \big( i \omega + \alf |\gamma|^2 \big)\, \dt + i \real \big\{ i \gamma \df \beta_t \big\} \Big)\, \wt{\q},
\eeq
\eeqs
where the second term on the right-hand-side (RHS) comes from the quadratic variation of the random phase.

Injecting next the stochastic ansatz \eqref{eqs:ansatz-const} as well as the noise definition \eqref{eq:noise-const} into the linearized system \eqref{eqs:rswlu-linear}, we obtain a system composed of differentiable terms and Brownian (martingale) terms that can be compactly written as
\beqs
\beq\label{eq:L-const}
\bL \wh{\q}_0\, \dt = 0,\ \quad \bL = \Big( -i \omega - \alf |\gamma|^2 + \alf \alpha^2 (\bk_\sigma \times \bk)^2 \Big)\, \bI_3 +
\begin{bmatrix} 
0 & -f_0 & i g k \\ 
f_0 & 0 & i g \ell \\ 
i H k & i H \ell & 0 
\end{bmatrix},     
\eeq
\beq\label{eq:Lp-const}
-i \real \big\{ i \gamma \df \beta_t \big\} + i \bk \bdot \real \big\{ i \alpha \bk_\sigma^\perp \df \beta_t \big\} = 0.
\eeq
\eeqs
The last equation leads to
\beq\label{eq:gamma-const}
\gamma = \alpha \bk_\sigma \times \bk \in \mb{R}.
\eeq
Substituting it into \eqref{eq:L-const}, we deduce
\beqs
\beq
\bL = 
\begin{bmatrix} -
i \omega & -f_0 & i g k \\ 
f_0 & -i \omega & i g \ell \\ 
i H k & i H \ell & -i \omega  
\end{bmatrix}.   
\eeq
We remark that in this case the linear operator $\bL$ for the resolved variables reduces to the same as that of the deterministic system \cite{Majda2003,Vallis-17}. 
Solving subsequently $\det (\bL) = 0$ gives us
\beq\label{eq:omega-const}
\omega = 0,\ \quad \omega = \pm \sqrt{g H |\bk|^2 + f_0^2}.
\eeq
\eeqs
We next construct the stochastic plane wave solutions associated to these frequencies.

\subsubsection{Stochastic Poincar\'e waves}

With the value of the last two frequencies in \eqref{eq:omega-const} and the associated eigenvector of $\bL$, one can find the following polarization relations:
\beq
\wt{\q} (\x, t) = 
\begin{bmatrix} 
\frac{\omega k + i f_0 \ell}{H |\bk|^2} \\ 
\frac{\omega \ell - i f_0 k}{H |\bk|^2} \\ 
1 
\end{bmatrix} 
\wh{\eta}_0 \exp \Big( i \big( \bk \bdot \x - \omega t + \gamma \imag \{ \beta_t \} \big) \Big).
\eeq  
Taking the real part of this ansatz, we deduce the path-wise wave solution:
\beqs\label{eqs:wave-const}
\beq
\eta (\x, t) = \wh{\eta}_0 \cos \big( \bk \bdot \x - \omega t + \gamma \imag \{ \beta_t \} \big) ,
\eeq 
\begin{align}
\bu (\x, t) &= \frac{\wh{\eta}_0 \omega}{H |\bk|} \cos \big( \bk \bdot \x - \omega t + \gamma \imag \{ \beta_t \} \big) \frac{\bk}{|\bk|}  \nonumber \\
&+ \frac{\wh{\eta}_0 f_0}{H |\bk|} \sin \big( \bk \bdot \x - \omega t + \gamma \imag \{ \beta_t \} \big) \frac{\bk^\perp}{|\bk|} .
\end{align}
\eeqs
In this simple case, we can analytically compute the ensemble-mean from these path-wise wave solutions. Note that $X_t := \bk \bdot \x - \omega t + \gamma \imag \{\beta_t\} \sim \mc{N} (\bk \bdot \x - \omega t, \gamma^2 t)$, and the characteristic function of the Gaussian process $X_t$ is given by $\Exp \big[ \exp (i X_t) \big] = \exp \big( i (\bk \bdot \x - \omega t) - \alf \gamma^2 t \big)$. 
One can then deduce the mean of the random ansatz \eqref{eqs:ansatz-const}, taking  its real part leads finally to
\beqs
\beq    
\Exp \big[ \eta \big] = \wh{\eta}_0 \cos \big( \bk \bdot \x - \omega t \big) \exp \big( -\alf \gamma^2 t \big),
\eeq    
\beq
\Exp \big[ \bu \big] = \frac{\wh{\eta}_0}{H |\bk|^2} \Big( \omega \bk \cos \big( \bk \bdot \x - \omega t \big) + f_0 \bk^\perp \sin \big( \bk \bdot \x - \omega t \big) \Big) \exp \big( -\alf \gamma^2 t \big).
\eeq
\eeqs
It can be readily observed from the random dispersion relation \eqref{eq:gamma-const}, that we recover the general mean solution \eqref{eqs:mean-wave} presented in the previous section.

\subsubsection{Stochastic geostrophic mode}

The polarization for the eigenvalue $\omega = 0$ reads
\beq
\wt{\q} (\x, t) = 
\begin{bmatrix} 
-i \frac{g}{f_0} \ell \\ 
 i \frac{g}{f_0} k \\ 
1 
\end{bmatrix} 
\wh{\eta}_0 \exp \Big( i \big( \bk \bdot \x + \gamma \imag \{ \beta_t \} \big) \Big).
\eeq  
The path-wise wave solutions are then given by
\beq
\eta = f_0 \wh{\eta}_0 \cos \big( \bk \bdot \x + \gamma \imag \{ \beta_t \} \big),\ \quad \bu = - g \wh{\eta}_0  \bk^\perp \sin \big( \bk \bdot \x + \gamma \imag \{ \beta_t \} \big).
\eeq
Similarly, one can recover the general mean solution \eqref{eqs:steady-mean} by taking the expectation of these path-wise solutions. As in the previous case only the waves that are excited by the stochastic forcing remain active on long terms horizon. 


\subsection{Approximation of path-wise waves under homogeneous noise}

We now extend the previous solution to statistically homogeneous noise. In the same way as previously we will assume a monochromatic noise as defined in \eqref{eq:a-const}, 
but now slowly varying in space: 
\beq\label{eq:noise-homog-approx}
\noi = \real \big\{ i \alpha \bks^\perp \exp (i \epsilon \bks \bdot \x) \df \beta_t \big\},\ \quad \ba = |\alpha|^2 \bks^\perp (\bks^\perp)\tp,
\eeq
where $\epsilon \ll 1$ is a small parameter to ensure the smooth structure of the noise. 
To infer wave solutions for such homogeneous noise, we now look for the following ansatz:
\beqs
\beq
\wt{\q}\, (\x, t) = \wh{\q}_0\, \exp \Big( i \big( \bk \bdot \x - \omega t - \real \big\{ i \gamma \exp (i \epsilon \bks \bdot \x) \beta_t \big\} \big) \Big),
\eeq
which generalizes our previous ansatz to homogeneous noise.
Applying the It\^o formula for this ansatz, we have
\beq
\df_t \wt{\q} = - \Big( \big( i \omega + \alf |\gamma|^2 \big)\, \dt + i \real \big\{ i \gamma \exp (i \epsilon \bks \bdot \x) \df \beta_t \big\} \Big)\, \wt{\q}.
\eeq
\eeqs
Injecting these solutions ansatz into system~\eqref{eqs:rswlu-linear}, we can separate again the Brownian parts and the time-differentiable component. The former reads
\beqs
\beq\label{eq:Lp-homog}
- i \real \big\{ i \gamma \exp (i \epsilon \bks \bdot \x) \df \beta_t \big\} + i \bk \bdot \real \big\{ i \alpha \bks^\perp \exp (i \epsilon \bks \bdot \x) \df \beta_t \big\} = 0 ,
\eeq
which leads to
\beq\label{eq:gamma-homog}
\gamma = \alpha \bk_\sigma \times \bk \in \mb{R}.
\eeq
\eeqs
In a similar way to the previous case, substituting this random dispersion into the diagonal component of the resolved linear operator satisfying $\bL \wh{\q}_0\, \dt = 0$, we get $\text{diag} (\bL) = -i \omega \bI_3$. However, in order to compute the gradient terms of the anti-symmetric part of $\bL$, the random phase is linearized as
\begin{align}
\wt{\q}\, (\x, t) &\approx \wh{\q}_0\, \exp \Big( i \big( \bk \bdot \x - \omega t - \gamma \real \big\{ i (1 + i \epsilon \bks \bdot \x) \beta_t \big\} \big) \Big), \nonumber\\
&= \wh{\q}_0\, \exp \Big( i \big( (\underbrace{\bk +  \epsilon \bk_\sigma \gamma \real \{ \beta_t \}}_{:=\wt{\bk}_t}) \bdot \x - \omega t + \gamma \imag \{ \beta_t \} \big) \Big).
\end{align}
Hereafter, $\wt{\bk}_t = [\wt{k}, \wt{\ell}]\tp$ is referred to as the effective wavenumber vector ensuing from the space varying random phase. It can be noticed that the real component of the complex Brownian path influences the wave's spatial phase, whereas the wave's temporal phase is randomized by the imaginary component. This latter has already been considered in the constant noise case. With the previous approximation, the linear operator $\bL$ can be finally written as
\beqs
\beq
\bL = 
\begin{bmatrix} -
i \omega & -f_0 & i g \wt{k}_t \\ 
f_0 & -i \omega & i g \wt{\ell}_t \\ 
i H \wt{k}_t & i H \wt{\ell}_t & -i \omega  
\end{bmatrix}.
\eeq
The two resulting dispersion relations are now given by
\beq\label{eq:omega-homog}
\omega = 0,\ \quad \omega = \pm \sqrt{g H \big| \wt{\bk}_t \big|^2 + f_0^2}.
\eeq
\eeqs
The latter  random dispersion relation reduces to the previous relation \eqref{eq:omega-const} (associated to a spatially constant noise) when $\epsilon =0$. We note that the homogeneous random noise leads to wave scattering. Such phenomena corresponds to similar results found in the setting of the  Wentzel-Kramers-Brillouin approximation \cite{Kunze-JPO-85,Mooers-GDF-75} or more recently through Wigner transform \cite{Danioux-Vanneste-PRF-16,Kafiabad_Savva_Vanneste_JFM_2019,kafiabad_vanneste_young_2021}. The stochastic framework explored here leads nevertheless to simpler formal developments.  

In the same way as previously, we exhibit in the following the two types of stochastic waves associated to this spatially slowly varying homogeneous noise.

\subsubsection{Stochastic Poincar\'e waves}

The path-wise wave solution in this case can be written as
\beqs
\beq
\eta (\x, t) = \wh{\eta}_0 \cos \big( \wt{\bk}_t \bdot \x - \omega t + \gamma \imag \{ \beta_t \} \big) ,
\eeq 
\begin{align}
\bu (\x, t) &= \frac{\wh{\eta} \omega}{H \big| \wt{\bk}_t \big|} \cos \big( \wt{\bk}_t \bdot \x - \omega t + \gamma \imag \{ \beta_t \} \big) \frac{\wt{\bk}_t}{\big| \wt{\bk}_t \big|} \nonumber\\  
&+ \frac{\wh{\eta} f_0}{H \big| \wt{\bk}_t \big|} \sin \big( \wt{\bk}_t \bdot \x - \omega t + \gamma \imag \{ \beta_t \} \big) \frac{\wt{\bk}_t^\perp}{\big| \wt{\bk}_t \big|} .
\end{align}
\eeqs
It can be remarked that the surface elevation phase has two sources of randomness: a temporal one, in the modified frequency, and one in space coming from the space varying noise. 
For $\bk = \bk_\sigma$ the solutions corresponds again to the classical deterministic waves, while, as shown in section \ref{Sec-Mean}, when $\bk \neq \bk_\sigma$, the mean of the stochastic wave solutions \eqref{eqs:mean-wave} is damped compared to the deterministic surface elevation ($h_d (\x, t)$). 

\subsubsection{Stochastic geostrophic mode}

The path-wise wave solutions are given by
\beq
\eta = f_0 \wh{\eta}_0 \cos \big( \wt{\bk}_t \bdot \x +  \gamma \imag \{ \beta_t \} \big),\ \quad \bu = - g \wh{\eta}_0  \wt{\bk}_t^\perp \sin \big( \wt{\bk}_t \bdot \x + \gamma \imag \{ \beta_t \} \big).
\eeq
These solutions correspond to dispersive geostrophic modes wave packet. The velocity wave is a pure vorticity wave packet. Its ensemble mean is also damped  for $\bk \neq \bk_\sigma$ and corresponds to the geostrophic stationnary wave for $\bk = \bk_\sigma$ \eqref{eqs:steady-mean}.

As a final word, on the general linear stochastic shallow water system, it can be noticed that in the long wave limit with $|\wt{\bk}|^2 \ll 1 / L_d^2$, (when the frequency approaches the Coriolis frequency), the pressure gradient force becomes negligible compared to the other terms in \eqref{eq:moment-linear}, and for a noise amplitude of order unity the linear shallow water system boils down to a linear stochastic transport equation in a rotating frame : 
\beq
\df_t \bu + \Big( f_0 \bu^\perp - \alf \bdiv (\ba \grad \bu) \Big)\, \dt + \noi \adv \bu = 0.
\eeq
Up to the diffusion and noise term (whose energy balances exactly) this closely corresponds to the so-called near inertial regime in which the fluid is purely transported. 
In the LU setting, the noise acts always as a random deviation whose energy is exactly compensated by the diffusion term. In the particular case of the shallow water model (linear or nonlinear) the total energy is conserved.  For the linear system the total energy of the mean being damped up to a constant,  the total energy variance increases up to a constant as a consequence of the total energy conservation. 

\subsection{Numerical illustrations}

We next give simple illustrations of the stochastic wave solutions. Here, rather than evaluating the analytical solutions presented in the previous sections, we propose to discretize numerically the linearized RSW-LU system \eqref{eqs:rswlu-linear} and perform Monte-Carlo simulations in order to estimate both path-wise and ensemble-mean wave solutions. To that end, the spectral (Fourier) method is adopted for the spatial discretization within a periodic domain, and an exponential integrator \cite{Dinvay-Memin-22} combined with the Milstein scheme \cite{Fiorini-Stuod-22} is used to approximate the mild solution. This semi-discrete problem can be written as
\beqs
\beq\label{eq:q1}
\wh{\q^{\sub (1)}} = - i \bk \bdot \wh{(\q_t \bsig \Delta B_t)} ,
\eeq
\beq\label{eq:q2}
\wh{\q^{\sub (2)}} = - i \bk \bdot \wh{(\q^{\sub (1)} \bsig \Delta B_t)} ,
\eeq
\beq
\wh{\q}_{t + \Delta t} = \exp (\bs{A} \Delta t) \left( \wh{\q}_t + \wh{\q^{\sub (1)}} + \alf \wh{\q^{\sub (2)}} \right) ,
\eeq
\beq
\bs{A} = 
\begin{bmatrix} 
0 & f_0 & -i g k \\ 
-f_0 & 0 & -i g \ell \\ 
-i H k & -i H \ell & 0  
\end{bmatrix},\
\quad
\q = 
\begin{bmatrix}
u \\ v \\ \eta
\end{bmatrix},
\eeq
\eeqs
where $\wh{\bullet}$ denotes the projection coefficient on the discrete Fourier modes, $\q = \mc{F}^{-1} (\wh{\q})$ is the inverse dicrete Fourier transform of $\wh{\q}$, $\Delta t$ and $\Delta B_t$ stands for the timestep and the Brownian motion's increment respectively. We remark that the classical 2/3 dealiasing rule can be adopted for the practical computations of \eqref{eq:q1} and \eqref{eq:q2}.

A deterministic monochromatic wave corresponding to a single frequency of the Poincar\'e waves (propagating to the left) is fixed as the initial condition:
\beq
\wh{\q}_0 = 
\begin{bmatrix} 
\frac{\omega k + i f_0 \ell}{H |\bk|^2} \\ 
\frac{\omega \ell - i f_0 k}{H |\bk|^2} \\ 
1 
\end{bmatrix} 
\delta (\bk - \bk_0),\ \quad \omega = \sqrt{g H \big| \bk \big|^2 + f_0^2},
\eeq  
where $\delta$ denotes the Dirac function and $\bk_0$ is the initial wavenumber vector. 

For the simulation configuration, we consider a squared shallow basin of length $L = 5120$ km and depth $H = 100$ m at mid-latitude ($f_0 = 10^{-4}$ s$^{-1}$) and a large-scale wave with $\bk = [3 \Delta k, 0 \Delta \ell]\tp$, where $\Delta k = \Delta \ell = 2 \pi / L$. Each random system either under constant \eqref{eq:noise-const} or under homogeneous noises \eqref{eq:a-const} has been simulated over 5 years with 100 realizations. We remark that here we do not use the smoothness approximation \eqref{eq:noise-homog-approx} in the homogeneous case. The noise's amplitude is fixed as $\alpha = \sqrt{\tau} g/f_0$, where $\tau = \Delta x / \sqrt{g H}$ with $g = 9.81$ m s$^{-2}$, $\Delta x = 40$ km and $\Delta t = 5 \tau$.

\begin{figure}
\begin{center}
\includegraphics[width=4cm]{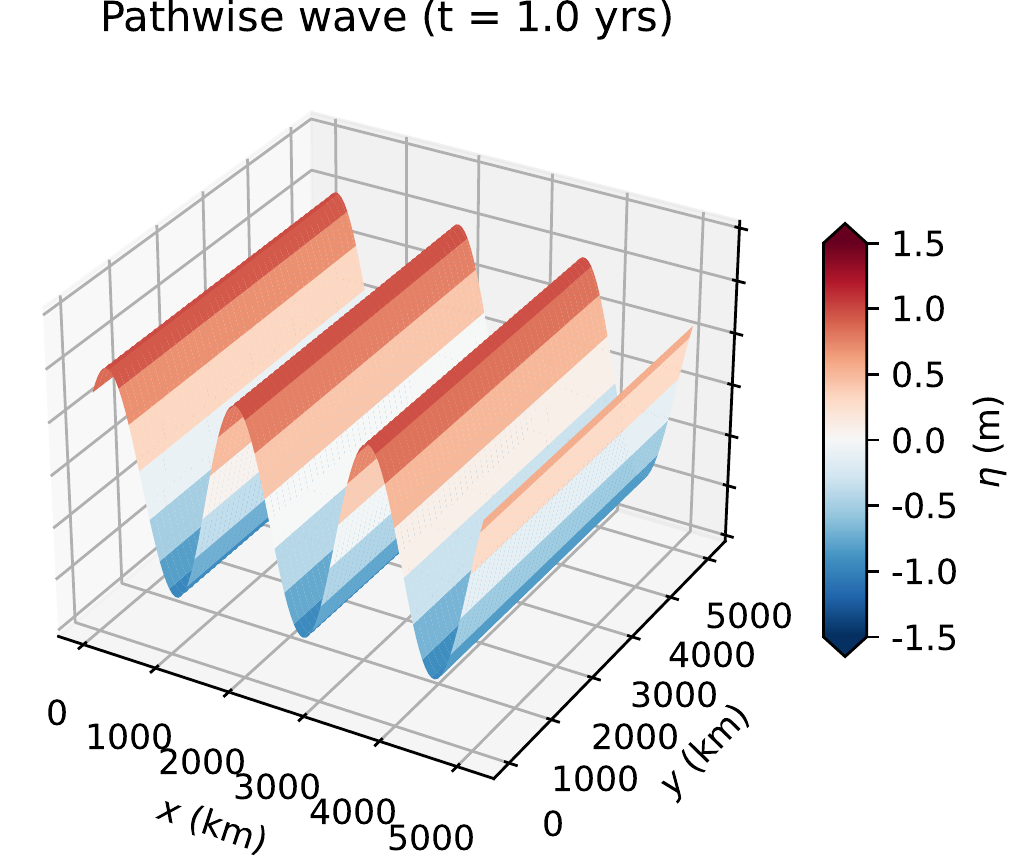}
\includegraphics[width=4cm]{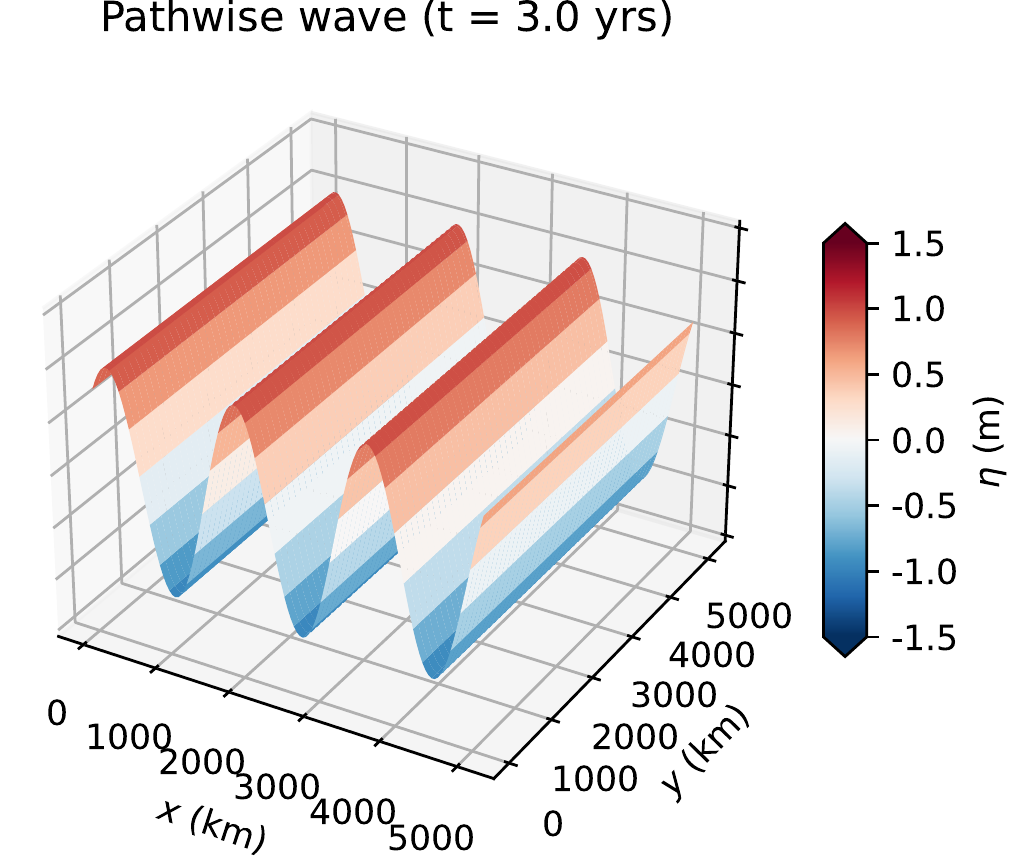}
\includegraphics[width=4cm]{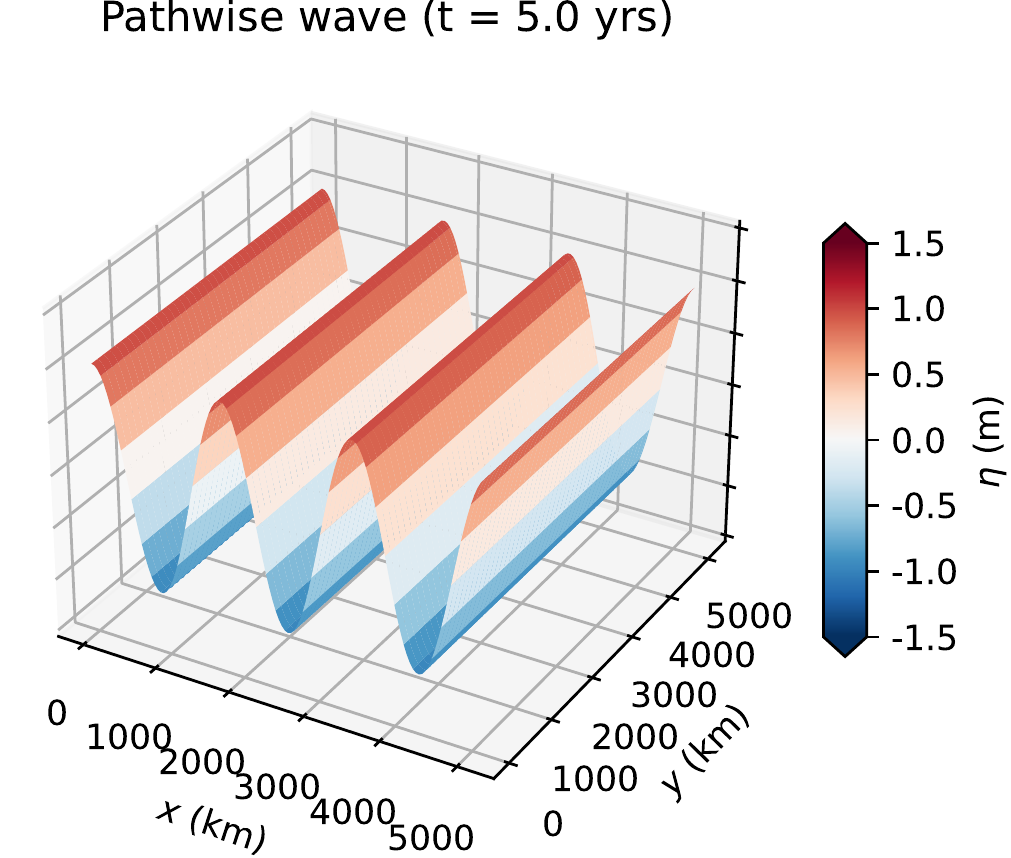} \\
\par\medskip
\includegraphics[width=4cm]{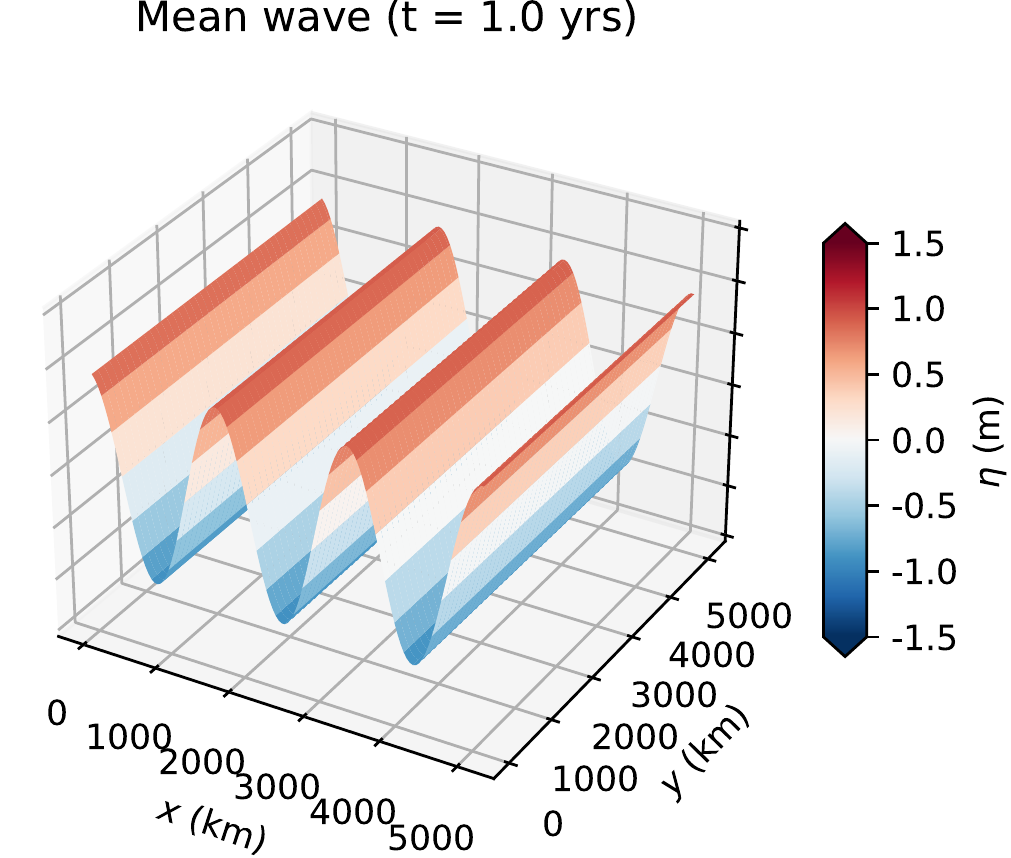}
\includegraphics[width=4cm]{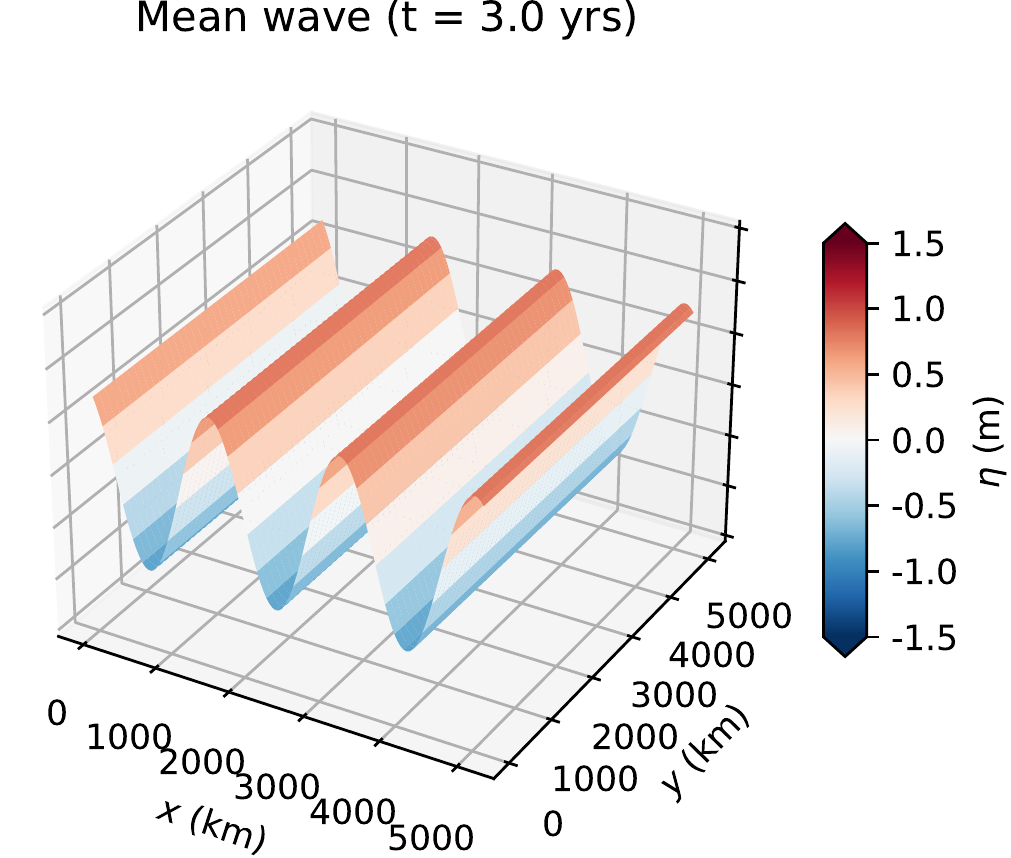}
\includegraphics[width=4cm]{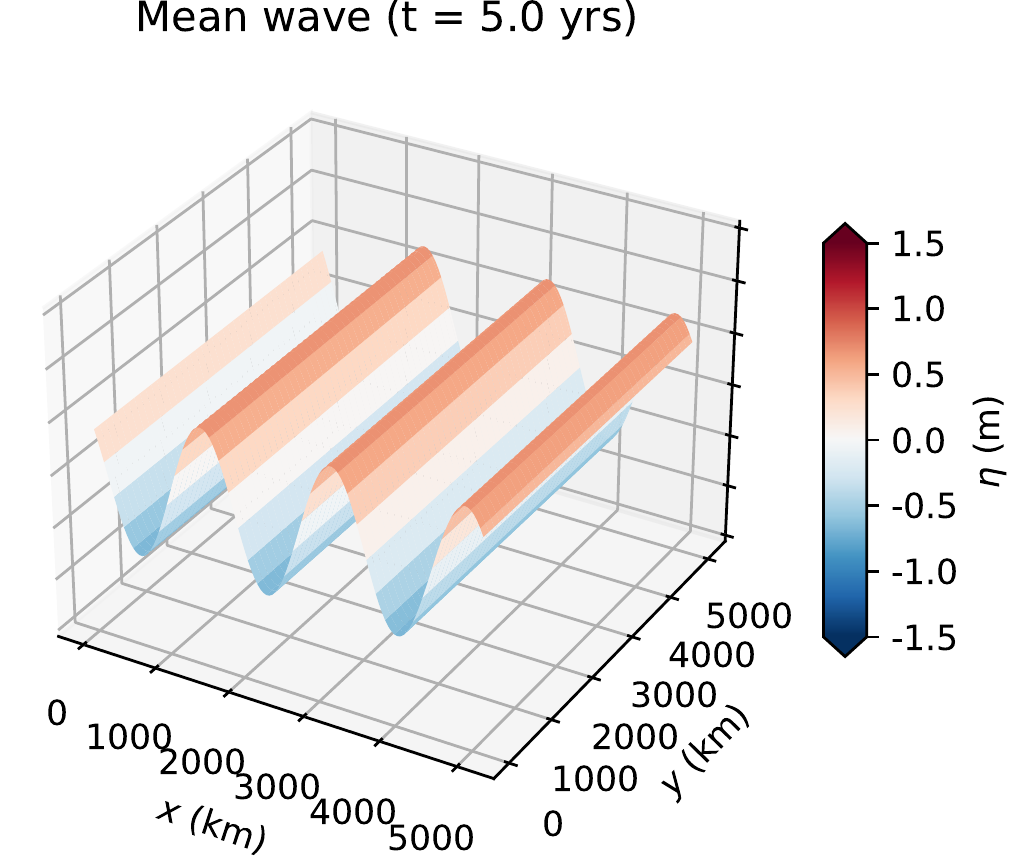}
\end{center}
\caption{Illustration for the path-wise surface elevation (top) and its ensemble-mean (bottom) with constant noise (with $\bks \times \bk \neq 0$) at different time (by columns).}
\label{fig:surf-const}
\end{figure}

Figure \ref{fig:surf-const} illustrates the evolution of the path-wise surface elevation $\eta$ and of the ensemble-mean  solution $\Exp [\eta]$ under a constant noise which has a different direction (with $\bks = [4 \Delta k, 6 \Delta \ell]\tp$) than that of the wave ($\bk_0$). In this case, 
the path-wise solution preserves the magnitude of the initial monochromatic wave while the ensemble-mean wave is damped along time. 

\begin{figure}
\begin{center}
\includegraphics[width=4cm]{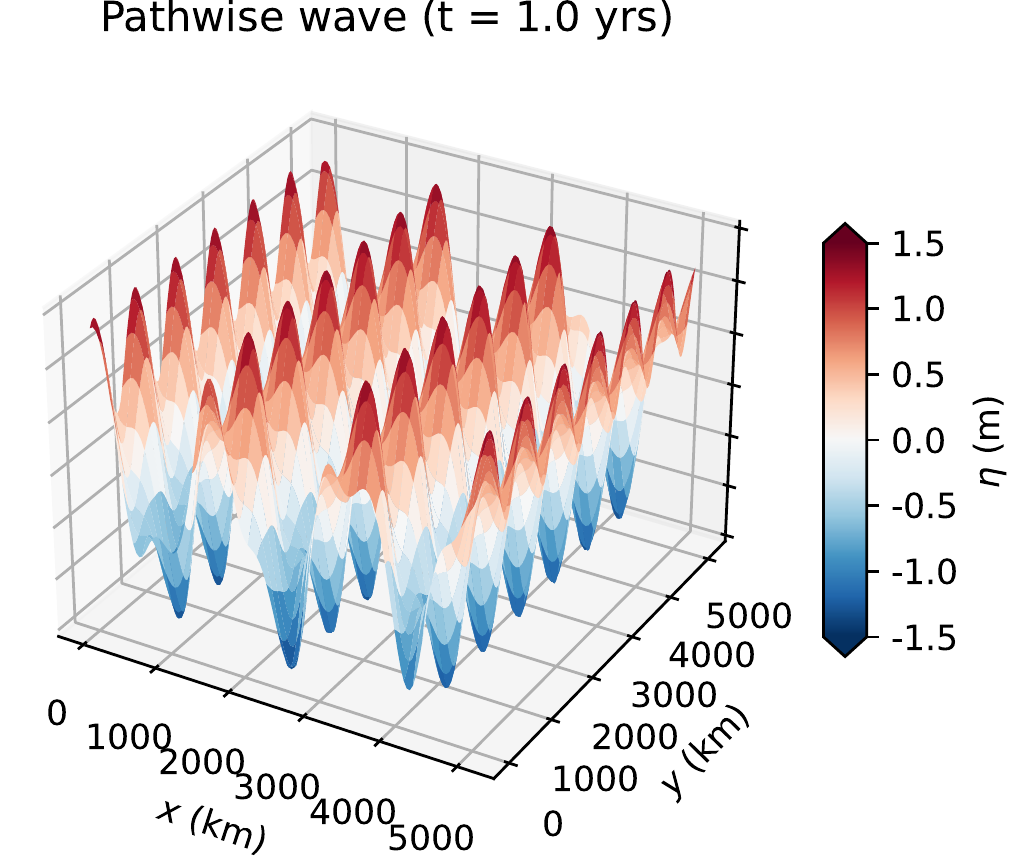}
\includegraphics[width=4cm]{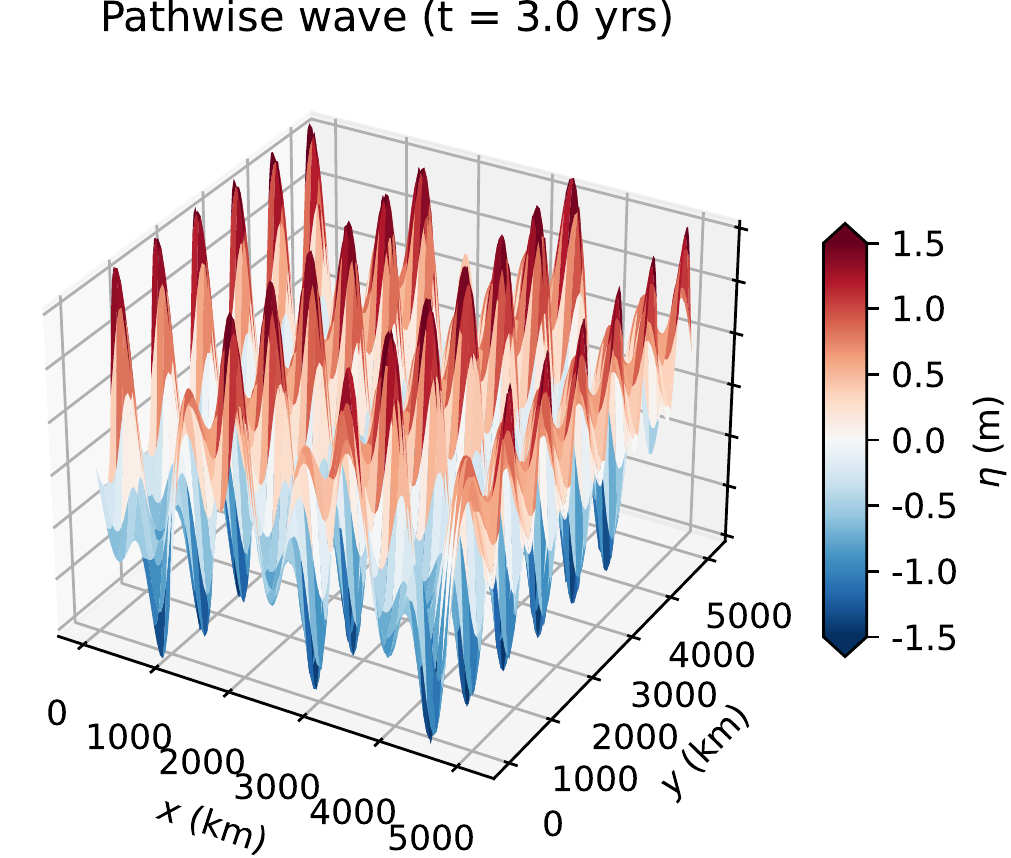}
\includegraphics[width=4cm]{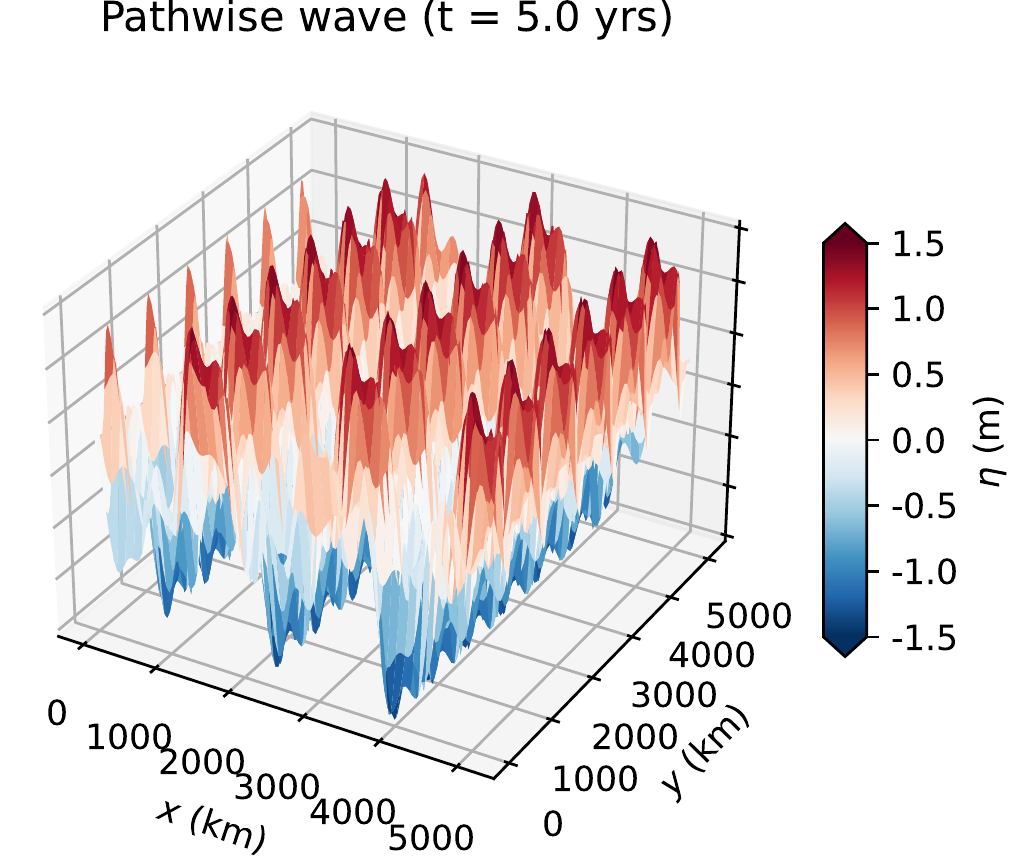} \\
\par\medskip
\includegraphics[width=4cm]{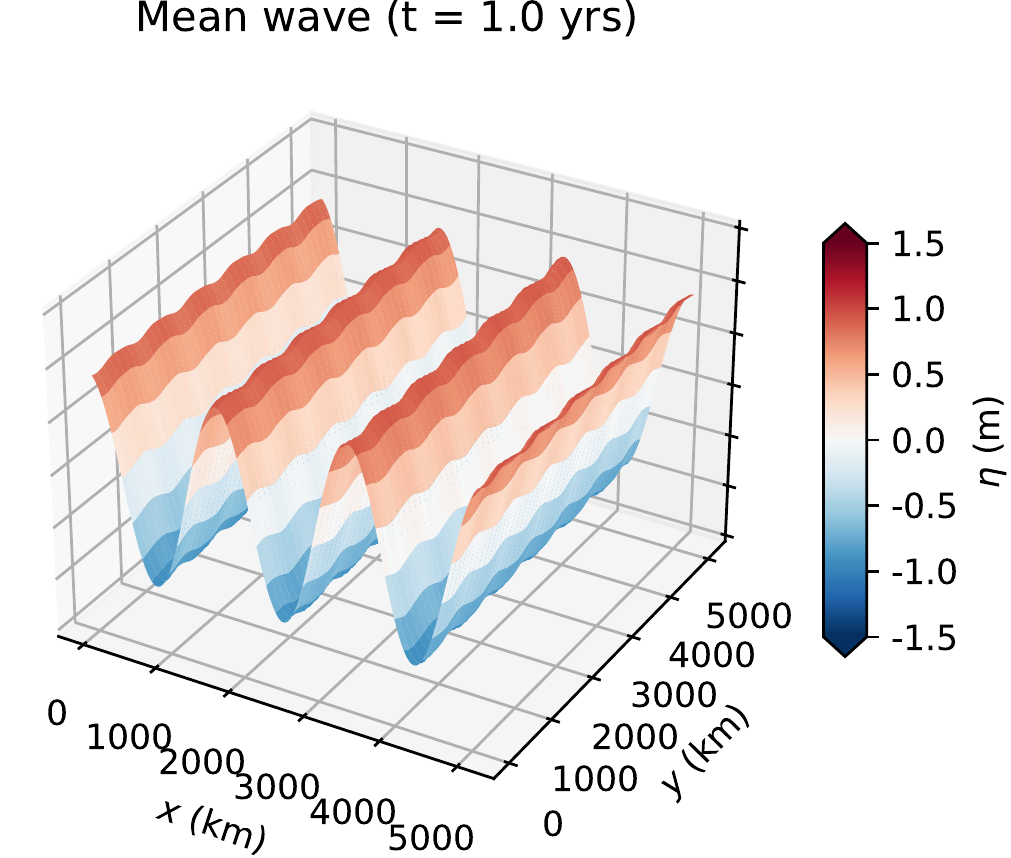}
\includegraphics[width=4cm]{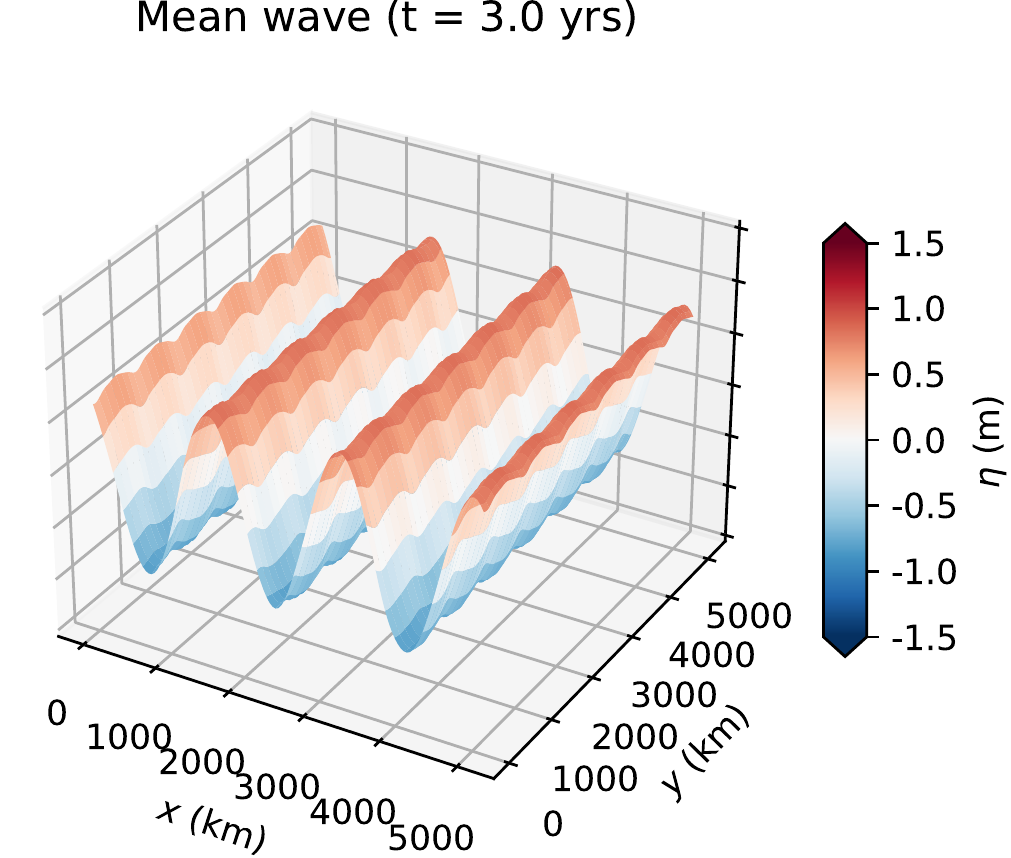}
\includegraphics[width=4cm]{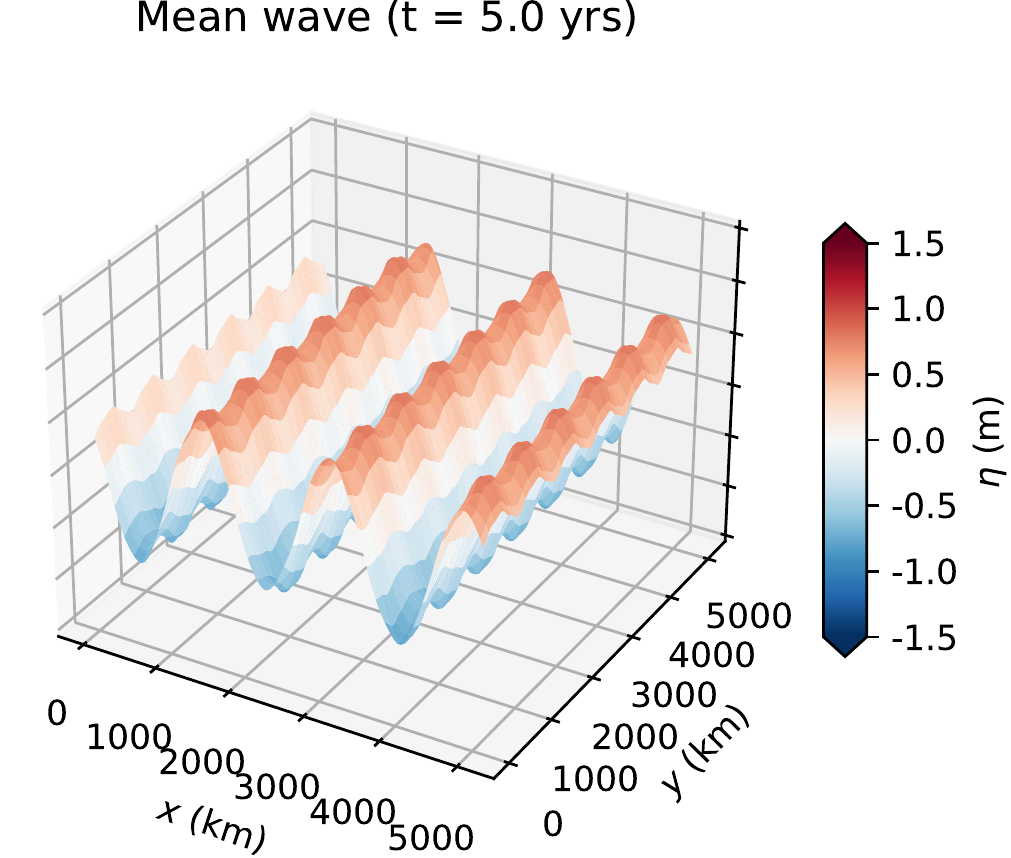}
\end{center}
\caption{Illustration for the path-wise surface elevation (top) and its ensemble-mean (bottom) with an homogeneous noise (with $\bks \times \bk \neq 0$) at different time (by columns).}
\label{fig:surf-homog}
\end{figure}

Figure \ref{fig:surf-homog} demonstrates the results obtained with the homogeneous noise (with 
the same $\bks$ as in the previous case). In that case, the path-wise wave is dispersive (scattering effect) whereas the mean wave is dissipative. 

The solutions related to the homogeneous noise is rougher than for a constant noise. The damping associated to the mean solution is visually similar in both case.

\begin{figure}
\begin{center}
\includegraphics[width=5.5cm]{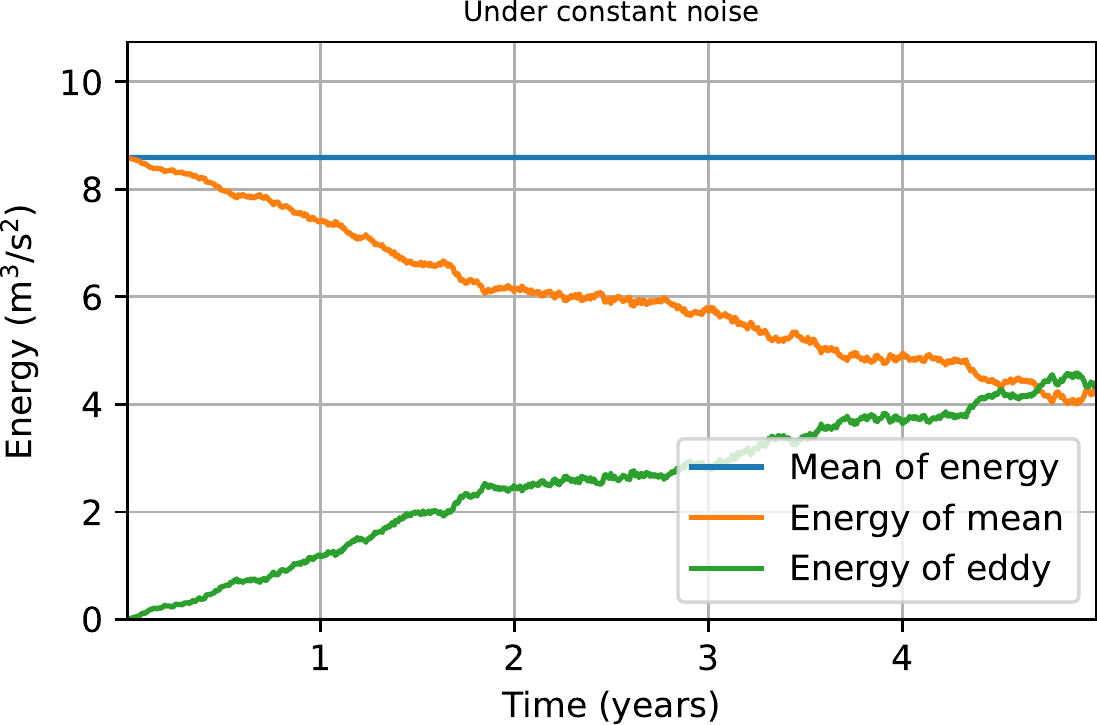}
\includegraphics[width=5.5cm]{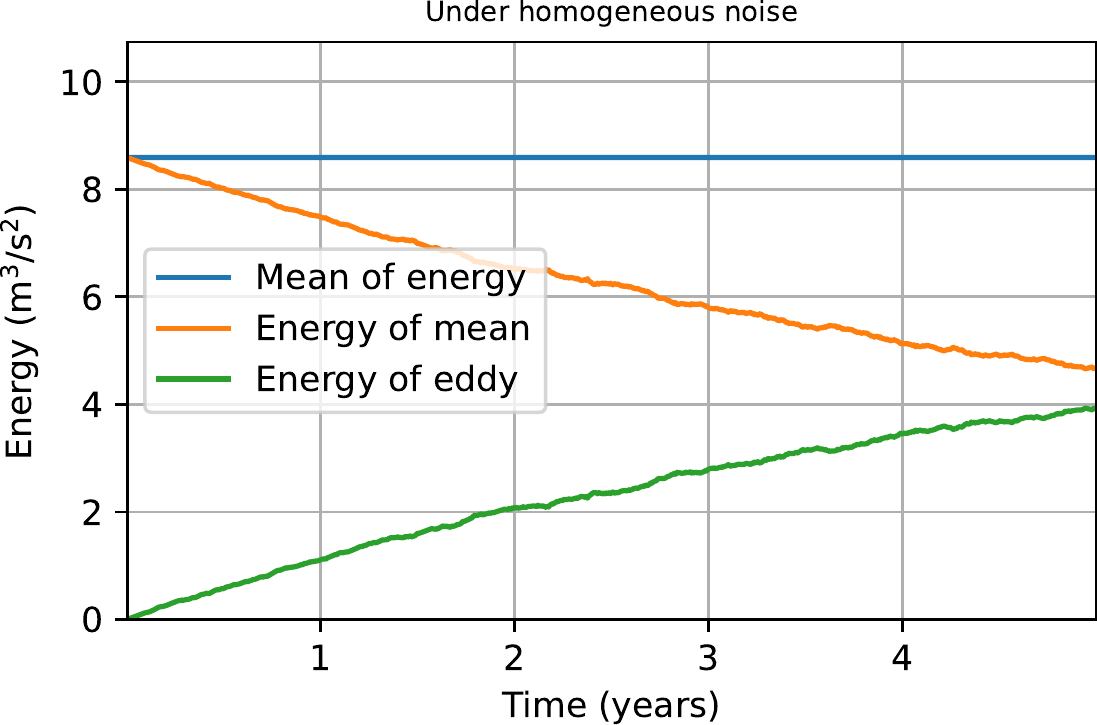} \\
\par\medskip
\includegraphics[width=5.5cm]{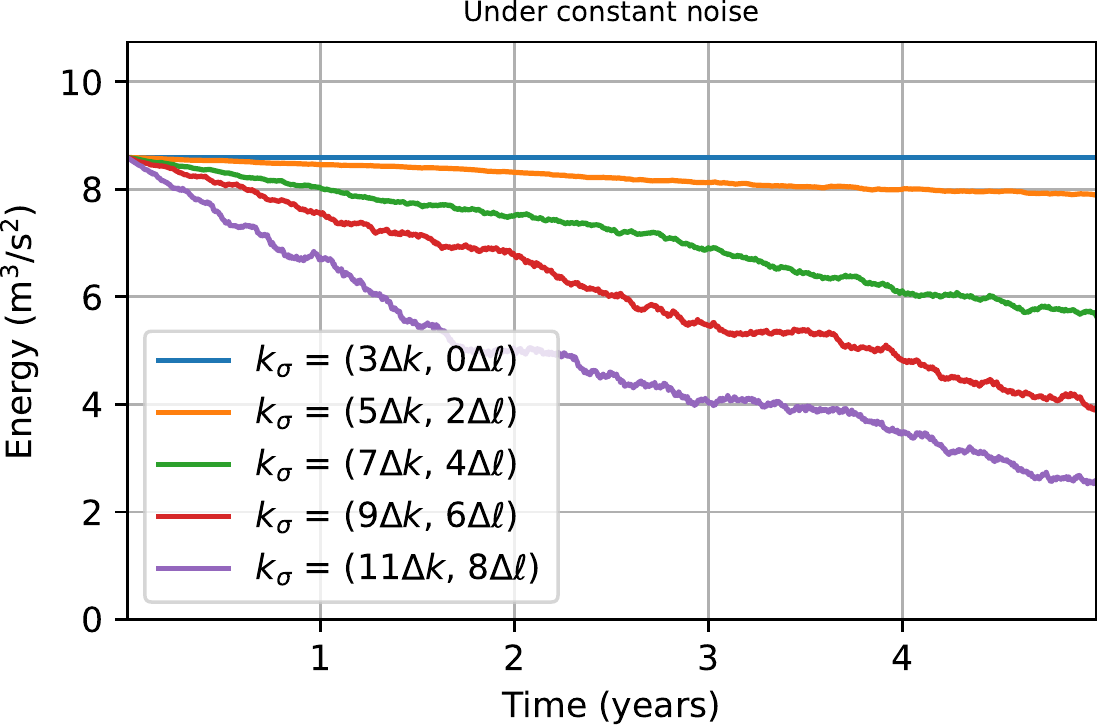}
\includegraphics[width=5.5cm]{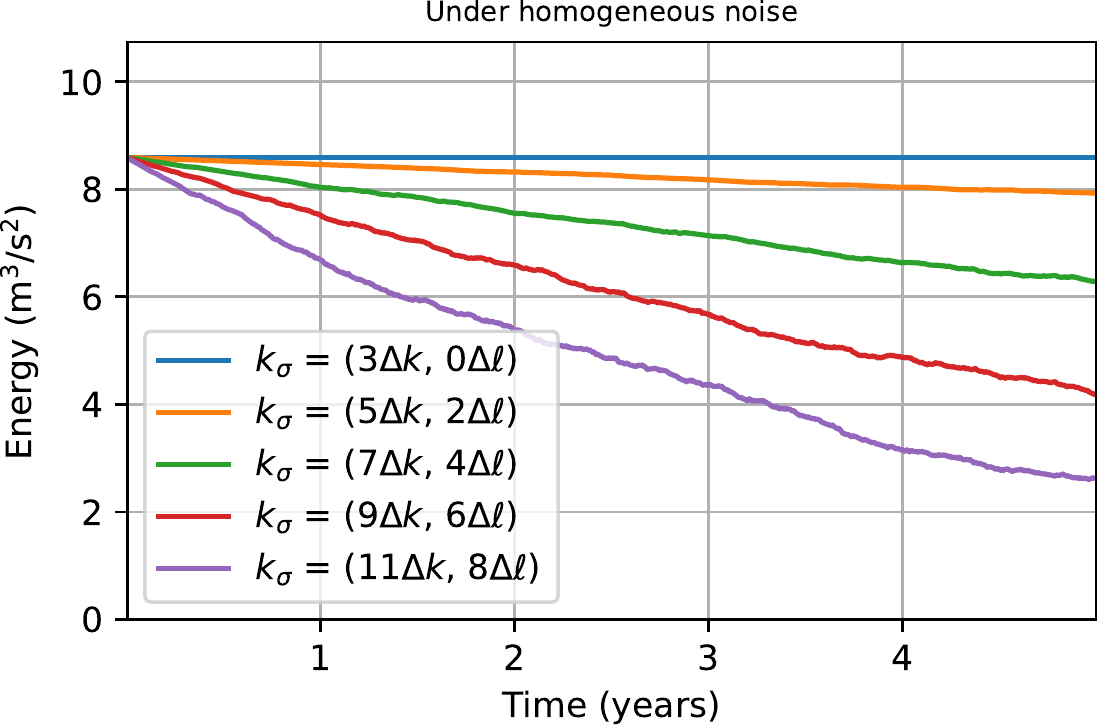} 
\end{center}
\caption{(Top) Time evolution of energy decomposition (w.r.t. ensemble) with constant (left) and homogeneous (right) noises; (Bottom) Comparison of dissipation of ensemble mean for different noise's scales.}
\label{fig:energy-diag}
\end{figure}

We then diagnosed an energy decomposition with respect to ensemble of different runs  in Figure \ref{fig:energy-diag}. This decomposition consists of the (ensemble) mean of energy $\Exp \big[ \int_{\dom} \alf (H |\bu|^2 + g \eta^2)\, \dx \big]$, the energy of (ensemble) mean $\int_{\dom} \alf \big( H |\Exp[\bu] |^2 + g \Exp[\eta]^2 \big)\, \dx$ and the energy of ``eddy" $\Exp \big[ \int_{\dom} \alf \big( H | \bu - \Exp[\bu] |^2 + g (\eta - \Exp[\eta])^2 \big)\, \dx \big]$. Figure \ref{fig:energy-diag} shows that for both noise models, the energy of mean waves is quickly dissipates along time while the energy of eddy waves increases with the same rate and continuously backscatters the variance to the ensemble. This mechanism ensures that the mean of the random energy is preserved in time. This corresponds well to the characteristic \eqref{eq:energy}, \eqref{total-energy-linear-sys} of the proposed stochastic transport model. 

Figure \ref{fig:energy-diag} illustrates also the dissipation rate of the energy of the mean waves in terms of different scales of the noise (equivalently, different angles of directions between the noise and the wave). 
The numerical results confirm our analyses in Section \ref{sec:sw-wave-lu}: 
the larger the angle $\theta$, or the smaller the noise's scale $\bks$, the faster the mean waves are damped in both cases.

\begin{figure}
\begin{center}
\includegraphics[width=4cm]{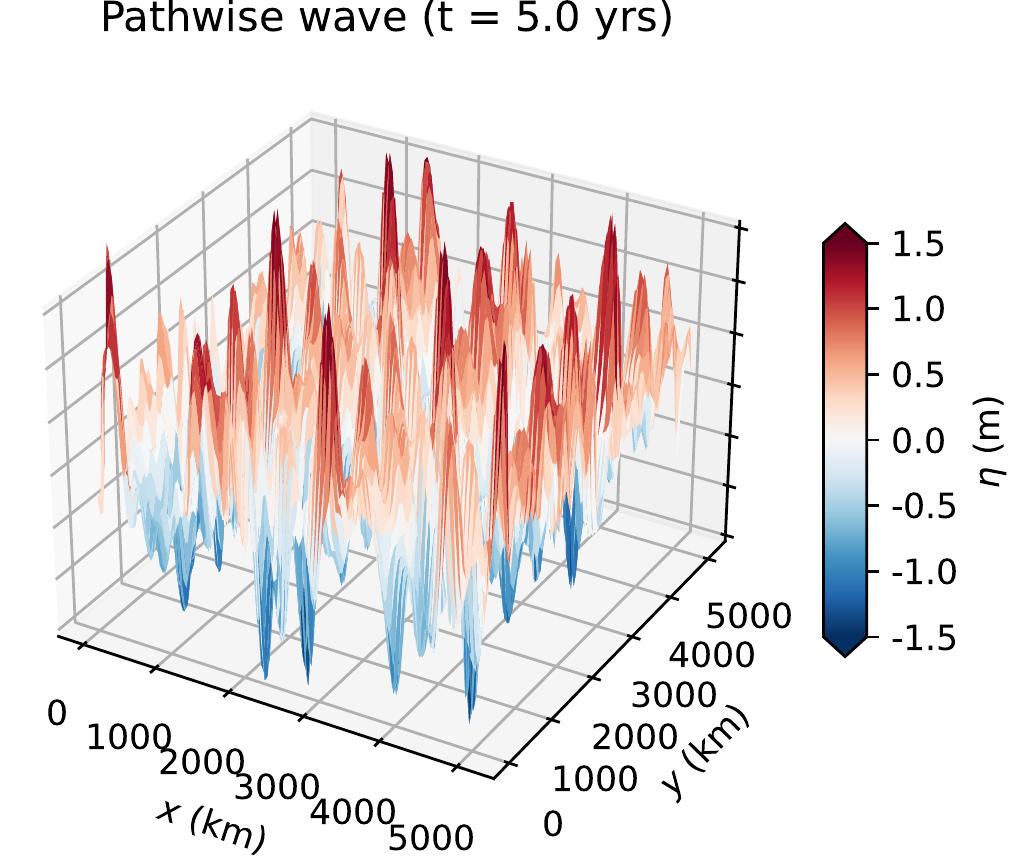} 
\includegraphics[width=4cm]{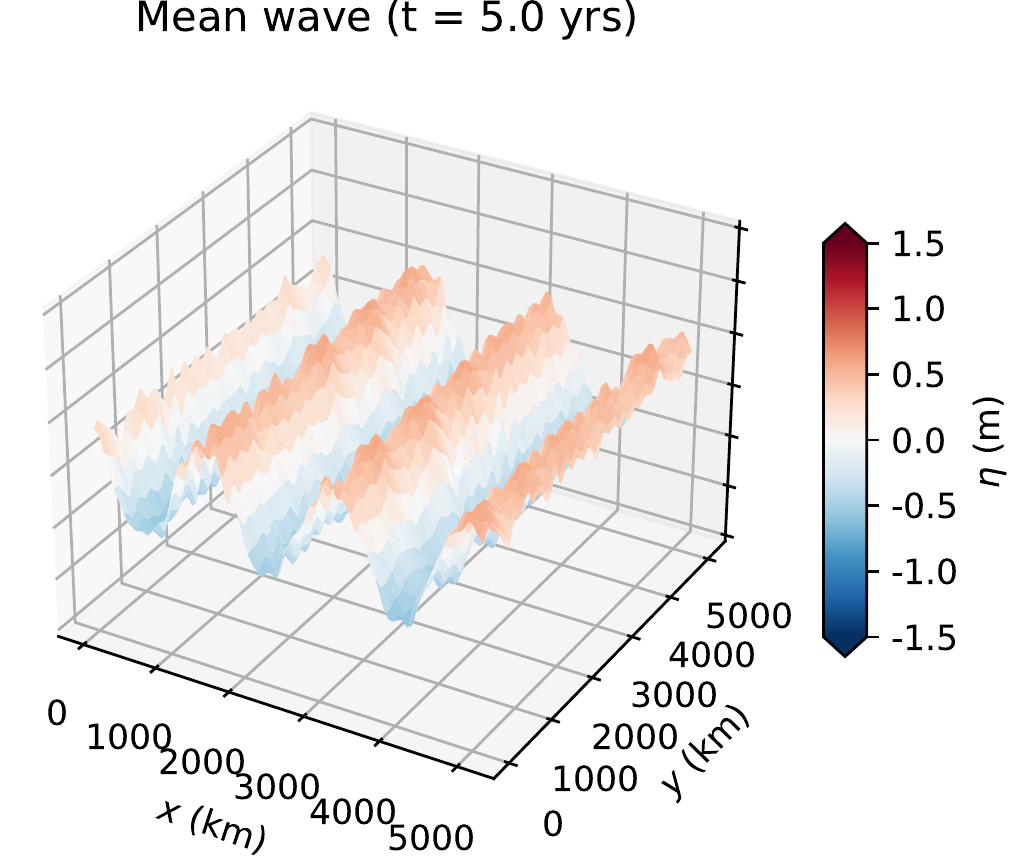} 
\includegraphics[width=4cm]{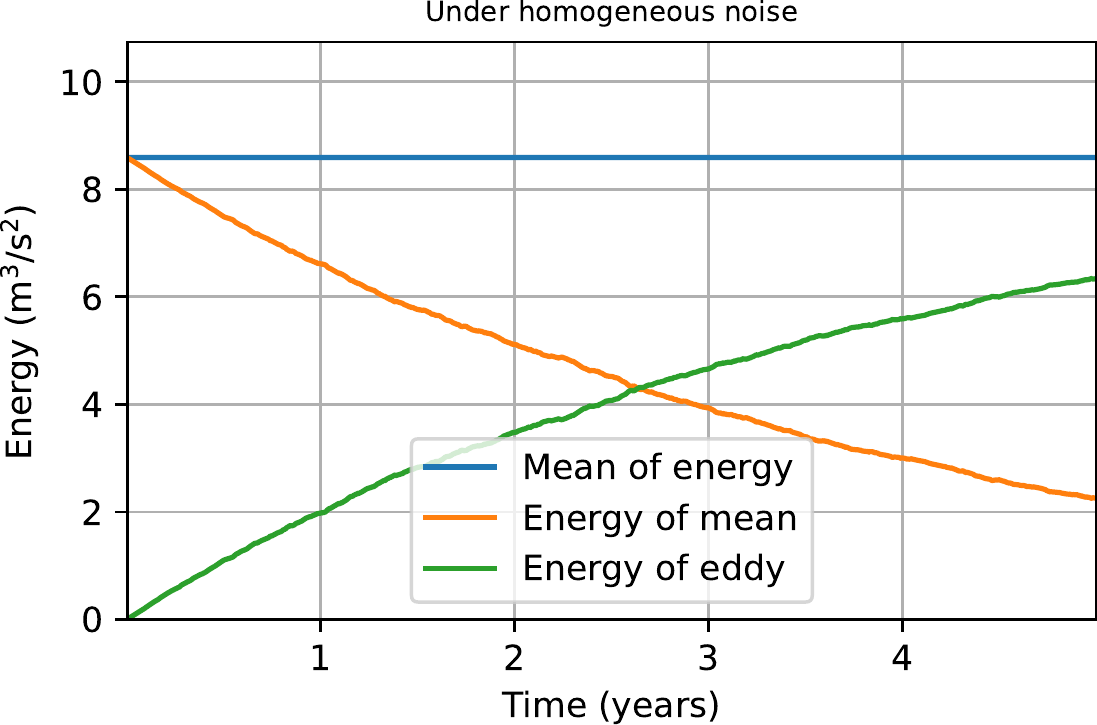}
\end{center}
\caption{Illustrations under multi-scale noise. From left to right: pathwise wave solution after 5 years simulation, the corresponding ensemble-mean and evolution of energy decomposition.}
\label{fig:multi-scale}
\end{figure}

Furthermore, propagation of the monochromatic wave by a muti-scale noise model have been numerically tested. In particular, we consider a band of wavenumbers for the noise, $\bks = \{\bk_j\}_{j=m,\ldots,M}$ with $|\bk_m| < |\bk_j| < |\bk_M|,\ \forall m<j<M$, satisfying the $-3$ spectrum power law: $\alpha_j^2 |\bk_j^\perp|^2 = \alpha_m^2 |\bk_m^\perp|^2\, (|\bk_j| / |\bk_m|)^{-3}$, i.e. $\alpha_j = \alpha_m (|\bk_j| / |\bk_m|)^{-5/2}$. Figure \ref{fig:multi-scale} shows the results with $\bk_m = [5 \Delta k, 5 \Delta \ell]\tp$, $\alpha_m = \alpha$ (as mentioned above) and 10 wavenumbers in total with uniform step $\Delta k$. After 5 years of simulation, the pathwise wave is dispersive while the mean wave is dissipative, and both of them are more irregular than the monochromatic noise solutions (see Figure \ref{fig:surf-homog}). We obtain also the consistent conclusion for the ensemble decomposition of the total energy. Moreover, the conversion from energy of mean to energy of eddy in this case is more faster and efficient than that of the monochromatic noise model (see Figure \ref{fig:energy-diag}).

\section{Shallow water PV dynamics and geostrophic adjustment}

The geostrophic adjustment is the process by which the flow and the pressure field tend to mutually adjust at large scale under the influence of  earth's rotation. An obvious manifestation of this adjustment corresponds to isobaric wind field and geostrophic current in the ocean. In the classical case of the deterministic linear shallow water model the geostrophic adjustment can be explained in terms of a variational formulation as the state of minimum energy corresponding to a given constant value of the potential vorticity (PV). Within the LU modelling, the geostrophic adjustment can be explained from the nonlinear system. To that end, let us derive the PV equation of the stochastic system \eqref{eqs:rswlu}. Taking first the curl of the momentum equation \eqref{eq:moment} under the space-time invariant Coriolis parameter $f_0$, we have 
\beqs\label{eqs:omega-h}
\begin{align}
\sto_t (\xi + f_0) &= - (\xi + f_0) \bdiv \bu\, \dt + \big[ \grad^\perp \bu, \grad \noi \big]_{\sub F} \nonumber \\
&+ \alf \sum_{\sub i,j=1,2} \pd_{x_i} \pd_{x_j} \big( \grad a_{ij} \times \bu \big)\, \dt,
\end{align}
where $\xi = \curl \bu = \pd_x v - \pd_y u$ denotes the relative vorticity and $[\bs{A}, \bs{B}]_{\sub F} = \mr{tr} (\bs{A}\tp \bs{B})$ stands for the Frobenius inner product of two matrices $\bs{A}$ and $\bs{B}$. 
Applying next the chain rule \cite{Resseguier-GAFD-I-17} for the stochastic transport of height \eqref{eq:mass}, one obtain
\beq
\sto_t h^{-1} = h^{-1} \bdiv \bu\, \dt.
\eeq
\eeqs
Let us recall the product rules of two stochastic transport equations in the following. In particular, if two arbitrary tracers $\theta$ and $\zeta$ satisfy
\beqs
\beq
\sto_t \theta = \Theta\, \dt,\ \quad \sto_t \zeta = Z\, \dt + \Sigma \df B_t,
\eeq
where $\Theta, Z$ are time-differentiable forcing terms and $\Sigma$ a martingale forcing component, then according to the It\^o-integration-by-part formula \cite{DaPrato}, we have
\beq
\sto_t (\theta \zeta) = \theta \sto_t \zeta + \zeta \sto_t \theta - \df \Big\langle \int_0^{t} \bs{\sigma}_s \df \B_s \adv \theta, \int_0^{t} \Sigma \df B_s \Big\rangle_t,
\eeq
\eeqs
where the last bracket term denotes the quadratic covariation of two martingales \cite{DaPrato}. Applying such rule for $\theta = h^{-1}$ and $\zeta = \xi + f_0$ associated with \eqref{eqs:omega-h}, one deduces the stochastic evolution of the PV, $q = (\xi + f_0) / h$, namely
\beqs
\begin{align}
\sto_t q &= h^{-1} \big[ \grad^\perp \bu, \grad \noi \big]_{\sub F} + \alf h^{-1} \sum_{\sub i,j=1,2} \pd_{x_i} \pd_{x_j} \big( \grad a_{ij} \times \bu \big)\, \dt \nonumber \\
&- \df \Big\langle \int_0^{t} \bs{\sigma}_s \df \B_s \adv h^{-1}, \int_0^{t} \big[ \grad^\perp \bu, \grad \bs{\sigma}_s \df \B_s \big]_{\sub F} \Big\rangle_t.
\end{align}
Opposite to the deterministic shallow water case the PV is not transported by the stochastic flow in general and some source/sink terms appear on the right-hand side of this PV equation. These source/sink terms reflect here the action of the small-scale on the non conservation of PV. In the deterministic context, PV is very sensitive to turbulence and subgrid modelling \cite{Bodner2020}. The same mechanism is at play here. We can nevertheless explore the condition for which PV remains conserved in the setting of a stochastic modeling of the small scales effect. The first and last terms cancel if the large-scale velocity field and the small-scale random component align with each other up to a uniform vector field. The second term trivially cancels if the random field is homogeneous in space (as in that case $\ba$ becomes a constant matrix). With these two previous conditions (alignment and homogeneity) PV is path-wise conserved. For homogeneous (incompressible) noise the expectation can be written in flux form and the mean PV is globally conserved. For homogeneous field, the PV equation reduces then to
\beq
\sto_t q = h^{-1} \big[ \grad^\perp \bu, \grad \noi \big]_{\sub F} - \df \Big\langle \int_0^{\bdot} \bsig \df \B_s \adv h^{-1}, \int_0^{\bdot} \big[ \grad^\perp \bu, \grad \bsig \df \B_s \big]_{\sub F} \Big\rangle_t.
\eeq
\eeqs
The above equation, combined with the previous results on the stochastic linear dynamics, enables revisiting the mechanism of geostrophic adjustment in the presence of a forcing. The corresponding global pictures is as follows.
Let us first consider random fluctuations in the ocean generated by wind forcing at large scales. Then, due to the wavelength mechanism described in the previous section, all the waves that are not aligned with this forcing are smoothed out and eventually annihilated following an exponential decay. The only waves that eventually remain are aligned with the wind forcing, which at large scale corresponds essentially to near inertial waves, and the PV dynamics corresponds then to a pure transport. Thus, the system tends then to relax to a balanced state as in the deterministic case. Conversely, we can devise a similar picture in the atmospheric context, in a configuration where the ocean plays the role of a noise for the atmosphere. The atmospheric waves will evolve toward a near inertial wave field aligned with the noise, following the exact same process. As a result, merging these two perspectives provides an interesting ocean/atmosphere coupling mechanisms of auto adjustment. At small scales, things are more complicated as the ISD has to be taken into account with an isotropization process that is likely less obvious. As a result the forcing terms in the PV equation remain. Furthermore, in that case the smooth spatial structure assumption of the noise imposes a very low noise amplitude. For the study of strong small-scale forcing this assumption as well as  the wave ansatz associated to it have to be revisited. This will be the objective of future works.


\section{Conclusions}

This stochastic extension of the shallow water equations highlights  the behavior of the LU setting for large-scale representation of flow dynamics. Opposite to classical eddy viscosity models, which introduces a similar damping term on the waves form, here, the waves that are sustained by the noise term keep their full classical expressions.  This provides a simple mechanism for waves selection and an interesting simplified model explaining the emergence of near inertial waves in ocean atmosphere systems. In the LU representation of the linearized shallow water system, the ensemble-mean waves that are not excited by a noise term with the same wavelength vanish exponentially fast whereas the others correspond to the classical deterministic wave solutions. The decay rate depends on the noise variance and on the wavenumber (in a quadratic way). The vanishing is therefore all the more fast for waves with small wavelength. The noise acts hence as a Dirac comb on the mean wave field.


\section*{Acknowledgments}

The authors acknowledge the support of the ERC EU project 856408-STUOD. The authors would like to thank Louis Thiry for his helpful comments, remarks and stimulating discussions. We also warmly thank Baylor Fox-Kemper and Oana Lang for fruitful discussions and advises on this work. The code to reproduce the numerical results is available at \url{https://github.com/matlong/sw-wave-lu}.

\bibliography{glob}

\end{document}